\documentclass[notitlepage,floats,aps,nofootinbib,preprintnumbers,superscriptaddress,twocolumn,prd,10pt,balancelastpage,longbibliography]{revtex4-1}

\usepackage{amsmath,amssymb,amsfonts}
\usepackage{mathtools}
\usepackage{hyperref}
\hypersetup{colorlinks,citecolor=blue,urlcolor=blue,linkcolor=blue}
\usepackage{graphicx}
\usepackage{xcolor}
\usepackage{nicefrac}
\usepackage{mathrsfs}
\usepackage{mdframed}
\usepackage{units}
\usepackage{slashed}

\newcommand{\gs}{g_\star}
\newcommand{\gss}{g_{\star s}}
\newcommand{\Trh}{T_\text{rh}}

\newcommand{\Tmax}{T_\text{max}}
\newcommand{\Hrh}{H_\text{rh}}
\newcommand{\arh}{a_\text{rh}}
\newcommand{\Mpl}{M_\text{Pl}}

\def\beq{\begin{equation}\begin{aligned}}
\def\eeq{\end{aligned}\end{equation}}

\begin{document}
\title{Next-to-Minimal Freeze-in Dark Matter}
\author{Nicolás Bernal}
\affiliation{New York University Abu Dhabi, PO Box 129188, Saadiyat Island, Abu Dhabi, UAE}
\author{Sagnik Mukherjee}
\author{James Unwin}
\affiliation{Department of Physics,  University of Illinois Chicago, Chicago, IL 60607, USA}

\begin{abstract} 
If the dark matter mass exceeds the highest temperature of the thermal bath, then dark matter production is Boltzmann suppressed. This opens new possibilities for dark matter model building. In particular, WIMP models that are experimentally excluded can be revived in this setting; conversely, freeze-in models, which would typically be beyond experimental reach, are potentially discoverable in the Boltzmann suppressed regime. In a recent letter, we highlighted these aspects for the case of electroweak doublet fermion dark matter assuming instantaneous inflationary reheating. Due to its elegance and simplicity, we coin this {\em Minimal Freeze-in} (MFI) Dark Matter. Here we consider next-to-minimal extensions of MFI dark matter. We present the implications for non-instantaneous reheating, including scenarios beyond the standard picture in which the Universe is initially matter dominated prior to reheating. Furthermore, we explore model variations within the electroweak dark matter scenario. Specifically, we consider fermion dark matter in higher representations of SU(2)${}_L$, exploring the current limits and the near-future discovery potential.  
\end{abstract}

\maketitle

\section{Introduction}

\vspace{-3mm}

There is an elegance in minimality, and models in which dark matter is connected to the Standard Model only via electroweak interactions epitomize this principle. Indeed, this is one of the original drivers of the popular WIMP scenario. In a companion paper, we laid out {\em Minimal Freeze-in} (MFI) dark matter  based on this notion \cite{Bernal:2026clv}. The particle content of MFI is a single pair of Weyl fermions that transform as doublets under SU(2)${}_L$. The neutral component of these doublets is the dark matter. It was highlighted that electroweak doublet dark matter is not excluded by direct detection (e.g.~LZ \cite{LZ:2024zvo}) if the dark matter mass $m$ is sufficiently large: $m\gtrsim10^{10}$~GeV.\footnote{If the Dirac fermion is split into distinct Majorana mass eigenstates by higher-dimensional operators, these limits are relaxed to 300~GeV, but this requires new high-scale physics \cite{Tucker-Smith:2001myb, Bernal:2026clv}.\label{fn1}} 
 
In addition to the requirements imposed by direct detection, to successfully realize MFI one must be in the Boltzmann suppressed freeze-in regime \cite{Giudice:2000ex, Cosme:2023xpa,  Cosme:2024ndc, Okada:2021uqk, Koivunen:2024vhr, Arcadi:2024wwg, Boddy:2024vgt, Arcadi:2024obp, Bernal:2024ndy, Bernal:2025qkj, Bernal:2026clv, Feiteira:2026qme}, with the dark matter mass exceeding the maximum temperature of the Universe: $m > \Tmax$. Notably, with this hierarchy the production rate typically never exceeds the Hubble rate and thus the dark matter never thermalizes. In the case where inflationary reheating is efficient, then $\Tmax = \Trh \simeq \sqrt{\Gamma\, \Mpl}$, where $\Gamma$ is the decay width of the inflaton. Notably, with this instantaneous reheating approximation MFI has only two free parameters, the dark matter mass $m$ and $\Trh$, which makes MFI both highly predictive and potentially discoverable at near-future experiments such as DARWIN \cite{DARWIN:2016hyl, Bernal:2026clv}. 

This paper explores extensions of the MFI scenario, thus {\em Next-to-Minimal Freeze-in dark matter}. We extend beyond the assumption of instantaneous inflationary reheating and consider variations on the original MFI particle content. On the cosmological side we explore non-instantaneous inflationary reheating, including scenarios in which the equation of state of the Universe is different from matter dominated immediately prior to reheating. Standard freeze-in calculations \cite{Hall:2009bx, Elahi:2014fsa} can be significantly impacted by departures from the instantaneous reheating approximation \cite{Garcia:2017tuj, Bernal:2019mhf, Garcia:2020eof, Bernal:2020bfj, Bernal:2020qyu, Allahverdi:2020bys, Garcia:2021iag, Barman:2022tzk, Batell:2024dsi, Bernal:2025fdr}. On the particle physics side we examine fermion dark matter in higher representations of SU(2)${}_L$, drawing inspiration from the `Minimal Dark Matter' scenario \cite{Cirelli:2005uq}. For higher representations, freeze-in is again mediated by the electroweak gauge bosons, but phenomenological differences arise. For dark matter with SU(2)${}_L$ representation of dimension $n\leq 9$ indirect detection limits on dark matter annihilation exclude the possibility that the relic density is set by thermal freeze-out \cite{Safdi:2025sfs}. These dark matter annihilation constraints only weakly constrain models of Boltzmann suppressed freeze-in since the dark matter is very much heavier \cite{Bernal:2026clv}, thus reviving these low $n$ scenarios. 

A major motivation of Minimal Dark Matter is that certain representations are cosmologically stable, without the need for {\em ad hoc} stabilizing symmetries. Interestingly, as dark matter can be heavier in Boltzmann suppressed freeze-in, the dark matter lifetime can be quite different. Certain scenarios retain the nice feature of dark matter metastability, whereas other models are constrained or excluded by searches for decaying dark matter.

This paper is structured as follows: In Section~\ref{sec2} we discuss the general coupling structure of arbitrary dimension representations of SU(2)${}_L$, and derive the relevant production cross-sections. We specialize to a number of models of particular interest in Section~\ref{sec4}. Working in the instantaneous reheating approximation, we identify the viable parameter space for dark matter being triplet ($\boldsymbol{3}$), quintuplet ($\boldsymbol{5}$), or septuplet ($\boldsymbol{7}$) under SU(2)${}_L$. Section~\ref{sec3} studies phenomenological aspects, including the dark matter relic abundance, cosmological bounds, direct and indirect detection searches. In Section~\ref{sec5} we explore the impact of deviating from instantaneous reheating, restricting our analysis to the doublet model. In Section~\ref{sec6}, we note the irremovable contribution from gravitational production, but show that it is always subleading. Section~\ref{sec7} provides some concluding remarks.

\section{Couplings and Cross-sections}
\label{sec2}
Two left-handed (LH) Weyl spinors that transform identically under the non-Abelian gauge groups, but with equal and opposite hypercharge, are referred to as a vector-like pair. Such vector-like pairs are interesting as they provide the minimal manner of adding non-trivial representations of the Standard Model gauge groups while avoiding gauge anomalies. Here we consider a vector-like fermion pair ($\chi_1,~\chi_2$) that transforms as $n$ dimensional representations of SU(2)${}_L$ and singlets under SU(3)${}_c$, with hypercharge $Y$, thus
\beq
    \chi_1=(\boldsymbol{1},\boldsymbol{n})_{Y} , \quad\quad \chi_2=(\boldsymbol{1},\boldsymbol{n})_{-Y} .
\eeq
The isospin of each is $T=(n-1)/2$ with their components labeled by $T_3 \in \{-T,  -T+1,  \dots,  T\} $. The electric charge of the state with third component isospin $T_3$ is  
\beq
    Q = T_3 + Y .
\eeq
After electroweak symmetry breaking, for each $Q$, one may form a Dirac fermion $\Psi_Q$ composed of matching electromagnetic (EM) charge components from $\chi_1$ and $\bar{\chi_2}$. Equivalently, for each isospin component $\mathcal{T}$ we identify the LH and RH components of the Dirac fermion as $\chi_1^{(\mathcal{T})}$ and $\overline{\chi}_2^{(-\mathcal{T})}$, respectively, so that $T_3^L = T_3^R := \mathcal{T}$. Accordingly,
$    Q\big[\chi_1^{(\mathcal{T})}\big]
    =
    Q\big[\overline{\chi}_2^{(-\mathcal{T})}\big]
    \equiv Q .$
A neutral state occurs iff $T_3=-Y$ for some allowed value of $T_3$. Explicit examples are given in Section~\ref{sec4}.

The couplings of the new fermions follow from their electroweak representations and hypercharge. For a Dirac fermion $\Psi$ with weak isospin $T_3^{L}$ and $T_3^{R}$ for its LH and RH components and electric charge $Q$, the vector and axial-vector couplings to the $Z$ are
\beq
    \label{qqq}
    g^{\Psi,Z}_{V,Q} &= \frac{e}{2 s_W c_W}\Big(T_3^{L} + T_3^{R} - 2 Q s_W^2\Big),\\
    g^{\Psi,Z}_{A,Q} &= \frac{e}{2 s_W c_W}\Big(T_3^{L} - T_3^{R}\Big),
\eeq
where $s_W\equiv \sin\theta_W$, $c_W\equiv \cos\theta_W$, and $e=g_2\, s_W$. The expressions we will derive shortly will be for a general vector mediator with the couplings corresponding to a Lagrangian of the form
\beq
    \mathcal L \supset Z_\mu \bar\Psi\gamma^\mu\big(g^{\Psi,Z}_{V,Q}-g^{\Psi,Z}_{A,Q}\gamma^5\big)\Psi.
\eeq
Substituting into Eq.~\eqref{qqq} gives, for the $\mathcal{T}^\text{th}$ component
\beq
    g^{\Psi,Z}_{V,Q}
    &= \frac{e}{s_W c_W}\Big(\mathcal{T}(1-s_W^2) - Y s_W^2\Big),\\
    g^{\Psi,Z}_{A,Q}
    &=0.
    \label{eq:couplings}
\eeq
Thus, for vector-like pairs the $Z$ coupling is purely vector,\footnote{Provided that the $\chi_i$ do not obtain an induced Majorana mass from higher-dimension operators, cf.~Refs.~\cite{Tucker-Smith:2001myb, Bernal:2026clv}.} whereas for the $W$ the couplings are
\beq
    g^{u_i d_j,W}_{V} &=g^{u_i d_j,W}_{A} =\frac{e}{2\sqrt{2}s_W}V_{ij}, \\
    g^{\Psi,W}_{V,Q} & =g^{\Psi,W}_{A,Q} = \frac{e}{2\sqrt{2}s_W}\sqrt{(T+\mathcal{T})(T-\mathcal{T}+1)},
\eeq
here, the $Q$ subscript in the coupling denotes the higher charge state in the co-production of a $\Psi_Q\Psi_{Q-1}$ pair.

We define $\beta(s) \equiv \sqrt{1-4(m^2/s)}$, then the cross-section for vector mediated production (as derived in  Appendix~\ref{ApA}) is 
\beq\label{xx}
   \sigma(q\bar q'\to \Psi_{Q}\bar\Psi_{Q'})
        \simeq \frac{\beta}{24\pi s}\mathcal{K}^{q\bar q'}_{QQ'},
\eeq
where $Q$, $Q'$ label the EM charges of the states, $q,q'=u_i,d_i$, and $\mathcal{K}^{q\bar q'}_{QQ'}$ is a combination of couplings.  We see in Eq.~\eqref{eq:couplings} that the axial $Z$ coupling is zero at tree-level for the vector-like pairs we consider, although the axial $W^\pm$ coupling is non-zero.

The specific form of $\mathcal{K}^{q\bar q'}_{QQ'}$ is given by
\beq \label{9}
    \mathcal{K}^{q\bar q'}_{QQ'} \equiv \delta_{qq'} \left(\mathcal{C}^{q\bar q}_{Q,\gamma} + \mathcal{C}^{q\bar q}_{Q,Z}\right) + (1 - \delta_{qq'})\mathcal{C}^{q\bar q'}_{QQ',W}
\eeq
with each term being a combination of couplings involving a different mediator  $i=\gamma,Z,W$,   defined by 
\beq\label{10}
    \mathcal C^{q\bar q'}_{Q,i} \equiv \Big( (g_{V}^{q\bar q',i})^2 +  (g_{A}^{q\bar q',i})^2\Big)  \Big((g^{\Psi,i}_{V,Q})^2+\beta^2(g^{\Psi,i}_{A,Q})^2\Big).
 \eeq
For the photon and $Z$ channel $q=q'$ and $Q=Q'$. At leading order in $\beta$ this is
\beq\label{11}
    \mathcal C^{q\bar q'}_{Q,i} \approx \Big( (g_{V}^{q\bar q',i})^2 +  (g_{A}^{q\bar q',i})^2\Big)  (g^{\Psi,i}_{V,Q})^2
\eeq
Note that $g^{\Psi,i}_{A,Q}=0$ for $i=\gamma,Z$,  whereas for $W$ mediation the axial term in $\mathcal C^{q\bar q'}_{Q,i}$ involves a $\mathcal{O}(\beta^2)$ suppression. This leading order limit is valid in the Boltzmann suppressed regime, since $m\gg \Trh$. 
For a given model, these couplings are determined by the SU(2)${}_L$ representation and the hypercharge of the vector-like pair. We provide numerical computations of $\mathcal C^{q\bar q'}_{Q,i}$ (as given in Eq.~\eqref{11}) in Tables~\ref{tab:CZ}-\ref{tab:CW} for a selection of  models of interest. Further exploration of these specific models is presented in Section~\ref{sec4} (the list of models is not exhaustive).

\begin{table*}[t]
\centering
\small
\renewcommand{\arraystretch}{0.95}
\caption{$\mathcal C_Z$}
\begin{tabular}{lccccc}
\hline\hline
Process
& Doublet 
& Triplet $(Y=1)$
& Triplet $(Y=0)$
& Quintuplet $(Y=0)$
& Septuplet $(Y=0)$
\\
\hline
$u_i\bar u_i\to \Psi^{+}\Psi^{-}$         & $1.59\times10^{-3}$ & $1.17\times10^{-3}$ & $1.30\times10^{-2}$ & $1.30\times10^{-2}$ & $1.30\times10^{-2}$ \\
$u_i\bar u_i\to \Psi^{++}\Psi^{--}$       & -   & $6.34\times10^{-3}$ & -   & $5.18\times10^{-2}$ & $5.18\times10^{-2}$ \\
$u_i\bar u_i\to \Psi^{+++}\Psi^{---}$     & -   & -   & -   & -   & $1.17\times10^{-1}$ \\
\hline
$d_i\bar d_i\to \Psi^{+}\Psi^{-}$         & $2.04\times10^{-3}$ & $1.51\times10^{-3}$ & $1.67\times10^{-2}$ & $1.67\times10^{-2}$ & $1.67\times10^{-2}$ \\
$d_i\bar d_i\to \Psi^{++}\Psi^{--}$       & -   & $8.17\times10^{-3}$ & -   & $6.68\times10^{-2}$ & $6.68\times10^{-2}$ \\
$d_i\bar d_i\to \Psi^{+++}\Psi^{---}$     & -   & -   & -   & -   & $1.50\times10^{-1}$ \\
\hline\hline
\end{tabular}
\label{tab:CZ}
%
\vspace{5mm}
\centering
\small
\renewcommand{\arraystretch}{0.95}
\caption{$\mathcal C_\gamma$}
\begin{tabular}{lccccc}
    \hline\hline
    Process
    & Doublet 
    & Triplet $(Y=1)$
    & Triplet $(Y=0)$
    & Quintuplet $(Y=0)$
    & Septuplet $(Y=0)$
    \\
    \hline
    $u_i\bar u_i\to \Psi^{+}\Psi^{-}$         & $4.28\times10^{-3}$ & $4.28\times10^{-3}$ & $4.28\times10^{-3}$ & $4.28\times10^{-3}$ & $4.28\times10^{-3}$ \\
    $u_i\bar u_i\to \Psi^{++}\Psi^{--}$       & -   & $1.71\times10^{-2}$ & -   & $1.71\times10^{-2}$ & $1.71\times10^{-2}$ \\
    $u_i\bar u_i\to \Psi^{+++}\Psi^{---}$     & -   & -   & -   & -   & $3.86\times10^{-2}$ \\
    \hline
    $d_i\bar d_i\to \Psi^{+}\Psi^{-}$         & $1.07\times10^{-3}$ & $1.07\times10^{-3}$ & $1.07\times10^{-3}$ & $1.07\times10^{-3}$ & $1.07\times10^{-3}$ \\
    $d_i\bar d_i\to \Psi^{++}\Psi^{--}$       & -   & $4.28\times10^{-3}$ & -   & $4.28\times10^{-3}$ & $4.28\times10^{-3}$ \\
    $d_i\bar d_i\to \Psi^{+++}\Psi^{---}$     & -   & -   & -   & -   & $9.64\times10^{-3}$ \\
    \hline\hline
\end{tabular}
\label{tab:Cgamma}
\vspace{5mm}
\centering
\small
\renewcommand{\arraystretch}{0.95}
\caption{$\mathcal C_W$}
\begin{tabular}{lccccc}
\hline\hline
Process
& Doublet
& Triplet $(Y=1)$
& Triplet $(Y=0)$
& Quintuplet $(Y=0)$
& Septuplet $(Y=0)$
\\
\hline
$u\bar d\to \Psi^{+}\bar\Psi^{0}$
& $5.35\times10^{-3}$ & $1.07\times10^{-2}$ & $1.07\times10^{-2}$ & $3.21\times10^{-2}$ & $6.43\times10^{-2}$ \\
$c\bar s\to \Psi^{+}\bar\Psi^{0}$
& $5.46\times10^{-3}$ & $1.09\times10^{-2}$ & $1.09\times10^{-2}$ & $3.28\times10^{-2}$ & $6.57\times10^{-2}$ \\
$t\bar b\to \Psi^{+}\bar\Psi^{0}$
& $5.69\times10^{-3}$ & $1.14\times10^{-2}$ & $1.14\times10^{-2}$ & $3.41\times10^{-2}$ & $6.84\times10^{-2}$ \\
\hline
$u\bar d\to \Psi^{++}\Psi^{-}$
& -   & $1.07\times10^{-2}$ & -  & $2.14\times10^{-2}$ & $5.35\times10^{-2}$ \\
$c\bar s\to \Psi^{++}\Psi^{-}$
& -   & $1.09\times10^{-2}$ & -   & $2.18\times10^{-2}$ & $5.46\times10^{-2}$ \\
$t\bar b\to \Psi^{++}\Psi^{-}$
& -   & $1.14\times10^{-2}$ & -   & $2.27\times10^{-2}$ & $5.69\times10^{-2}$ \\
\hline
$u\bar d\to \Psi^{+++}\Psi^{--}$
& -   & -   & -   & -   & $3.21\times10^{-2}$ \\
$c\bar s\to \Psi^{+++}\Psi^{--}$
& -   & -   & -   & -   & $3.28\times10^{-2}$ \\
$t\bar b\to \Psi^{+++}\Psi^{--}$
& -   & -   & -   & -   & $3.41\times10^{-2}$ \\
\hline\hline
\end{tabular}
\label{tab:CW}
\vspace{3mm}
\end{table*}

Notably, Eq.~\eqref{xx} can be adapted to different initial and final states
by taking the appropriate couplings, cf.~Eq.~\eqref{eq:couplings}. It is enlightening to consider two illustrative examples: the neutral current arising from the $\bar u u$ and the  charged current arising from the $\bar u d$ initial state
\begin{widetext}
    \beq\label{wide}
        \small \sigma \left(u\bar u \to \Psi_{Q}\bar\Psi_{Q}\right) &\simeq \frac{\beta}{24\pi s} \Bigg[ \underbrace{\Big(eQ_u\Big)^2 \Big(eQ\Big)^2}_{\mathcal C^{u}_{Q,\gamma}} +  \underbrace{\Big[\Big( \frac{e}{2s_W c_W}\left(\frac12-\frac{4}{3}s_W^2\right)\Big)^2+\Big(\frac{e}{4s_W c_W}\Big)^2\Big] \Big(\frac{e}{s_W c_W}\Big(\mathcal T-Qs_W^2\Big)\Big)^2}_{\mathcal C^{u}_{Q,Z}} \Bigg],\\
        \sigma(u\bar d\to \Psi_{Q}\bar\Psi_{Q-1}) &=  \frac{\beta}{24\pi s}\mathcal{C}^{u\bar d}_{Q,W} \simeq \frac{\beta}{24\pi s}\frac{e^4}{32 s_W^4}(T+\mathcal{T})(T-\mathcal{T}+1) |V_{ud}|^2.
    \eeq
\end{widetext}
Other cross-sections mediated by the $W$ bosons can be obtained by using the appropriate CKM element.

Further, we recall that if an EM neutral Dirac fermion exists (as appropriate to be dark matter), then $Q=0$ requires $Y =- \mathcal{T}$. This implies that for $\Psi^0$ the corresponding $Z$ vector coupling is
\beq
    g^{\Psi,Z}_{V,Q=0} = - \frac{e}{s_W c_W} Y .
    \label{eq:ZPsi0}
\eeq
It can be seen from Eq.~\eqref{eq:ZPsi0} that for integer isospin $T\in\mathbb Z$ (i.e.~odd $n$) and $Y=0$, the neutral state corresponds to $\mathcal{T}=0$ which satisfies $g^{\Psi^0,Z}_{V,0} = g^{\Psi^0,Z}_{A,0} = 0$, thus there is no diagonal tree-level coupling to the $Z$. The absence of tree-level $Z$-exchange significantly suppresses the direct detection cross-section which will only be generated at loop-level. In this case, the $Z$-mediated freeze-in cross-section for $\Psi^0$ also vanishes at leading order; however, the $W$-mediated piece persists at leading order and charged $\Psi$ states are also produced without suppression.

\section{Models of Interest}
\label{sec4}
We next outline explicitly a number of models of interest and then study the phenomenological implications. Specifically, we study the doublet ($\boldsymbol{2}$), the triplet ($\boldsymbol{3}$), the quintuplet ($\boldsymbol{5}$), and the septuplet ($\boldsymbol{7}$). We restrict our attention to fermion dark matter.

\subsection{Doublet Representations}
\label{doublet}
We start with the familiar case of the fundamental representation of SU(2)$_L$.  
For $n=2$ one has isospin $T=\frac{1}{2}$, and thus the components take values $\mathcal{T}\in\{-\frac{1}{2},+\frac{1}{2}\}$. For $Y=\pm\frac{1}{2}$ one of the components is EM neutral after electroweak symmetry breaking (EWSB). Taking $Y=\frac{1}{2}$ for definiteness, the electric charges are given by $Q_{\mathcal{T}}=\mathcal{T}+\frac{1}{2}\in\{0,+1\}$, with the EM neutral state corresponding to the $\mathcal{T}=-\frac{1}{2}$ component.

Labeling the two LH Weyl doublets by their hypercharge as $\chi^{(+1/2)}_{\mathcal{T}}$ and $\chi^{(-1/2)}_{\mathcal{T}}$. The associated Dirac fermions are then constructed as
\beq\notag
    \Psi^{0}\equiv
    \begin{pmatrix}
        \chi_{-1/2}^{(+1/2)}\\[8pt]
        \overline{\chi}_{+1/2}^{(-1/2)}
    \end{pmatrix},\qquad
    \Psi^{+}\equiv
    \begin{pmatrix}
        \chi_{+1/2}^{(+1/2)}\\[8pt]
        \overline{\chi}_{-1/2}^{(-1/2)}
    \end{pmatrix},
\eeq
together with $\Psi^{-}\equiv(\Psi^{+})^c$. The diagonal $Z$ couplings follow from Eq.~\eqref{eq:couplings}
\beq
    g^{\Psi,Z}_{V,Q} \Big|_{\shortstack{\footnotesize $n=2$\\ \footnotesize $Y=\frac12$}} =\frac{e}{s_W c_W}\Big(\mathcal{T}-(\mathcal{T}+Y)s_W^2\Big).
\eeq
Recall that in this case (and all cases studied below) the axial vector $Z$ coupling vanishes $g^{\Psi,Z}_{A,Q}=0$, due to the fact that we only consider vector-like pairs; see Eq.~\eqref{eq:couplings}.

For the neutral component (${\mathcal{T}}=-\frac{1}{2}$, $Q=0$) one has
\beq
    g^{\Psi,Z}_{V,0} \Big|_{\shortstack{\footnotesize $n=2$\\ \footnotesize $Y=\frac12$}} =- \frac{e}{2s_W c_W},
\eeq
while for the charged component 
\beq
    g^{\Psi,Z}_{V,+1} \Big|_{\shortstack{\footnotesize $n=2$\\ \footnotesize $Y=\frac12$}}=\frac{e}{2s_W c_W}\Big(1-2s_W^2\Big).
\eeq
Choosing instead $Y=-1/2$ simply relabels which $\mathcal{T}$ component is neutral and flips the overall sign of ${g^{\Psi,Z}_{V,0}}$; all production and scattering rates are unchanged since they depend only on $(g^{\Psi,Z}_{V,0})^2$.

The construction above is precisely the MFI model studied in the companion paper \cite{Bernal:2026clv}.  We present this as the point of comparison for the higher representations. Moreover, we will restrict our attention to this fermion doublet model in the non-minimal cosmology discussion presented in Sections \ref{sec5} \& \ref{sec6}.

\subsection{Triplet Representations}
The triplet $\boldsymbol{3}$ is the adjoint of SU(2)${}_L$. With $n=3$ one has isospin $T=1$, and thus the components take values $\mathcal{T}\in\{-1,0,+1\}$.  We shall first consider the hypercharge $Y=0$, followed by $Y=1$.  The case of $Y=0$ is distinct from $Y\neq0$ since the tree-level vector couplings vanish.

For $Y=0$, the electric charge is $Q=\mathcal{T}$ and a neutral component exists ($\mathcal{T}=0$). We label the two left-handed Weyl triplets $i=1,2$ by $\chi^{(1)}_{\mathcal{T}}$ and $\chi^{(2)}_{\mathcal{T}}$ (the hypercharges match so we replace this label), then the associated Dirac fermions for $Y=0$ are\footnote{For odd $n$ and $Y=0$ the theory is anomaly free with only a single Weyl fermion, the EW spectrum remains unchanged (up to multiplicity), and a Majorana mass is permitted. We retain the pair of $\chi$ fields here to maintain uniformity between cases. \label{fn3}} 
\beq\notag
    \Psi^{0}\equiv
    \begin{pmatrix}
        \chi^{(1)}_{0}\\
        \overline{\chi}^{(2)}_{0}
    \end{pmatrix},\quad
    \Psi^{+}\equiv
    \begin{pmatrix}
        \chi^{(1)}_{+1}\\
        \overline{\chi}^{(2)}_{-1}
    \end{pmatrix},\quad
    \Psi^{-}\equiv
    \begin{pmatrix}
        \chi^{(1)}_{-1}\\
        \overline{\chi}^{(2)}_{+1}
    \end{pmatrix}.
\eeq
The diagonal $Z$ couplings come from Eq.~\eqref{eq:couplings}
\beq\label{15}
    g^{\Psi,Z}_{V,Q} \Big|_{\shortstack{\footnotesize $n=3$\\ \footnotesize $Y=0$}} =\frac{e}{s_W c_W}\Big(\mathcal{T} - \mathcal{T} s_W^2\Big)=\frac{e c_W}{s_W} \mathcal{T}.
\eeq
Thus, the tree-level vector coupling is ${g^{\Psi,Z}_{V,Q}}^{(0)}=0$ for the neutral component. For $\Psi^\pm$ the couplings are
\beq
    {g^{\Psi,Z}_{V,\pm1}} \Big|_{\shortstack{\footnotesize $n=3$\\ \footnotesize $Y=0$}} =\pm \frac{e c_W}{s_W}.
\eeq

In contrast, in the $Y=1$ case, the EM charges are given by $Q=\mathcal{T}+1\in\{0,1,2\}$. Note that the neutral Dirac states $\Psi^0$ correspond to ${\mathcal{T}}=-1$. In this case, the set of Dirac fermions is formed as follows
\beq\notag
    \Psi^{0}\equiv
    \begin{pmatrix}
        \chi_{1,-1}\\
        \overline{\chi}_{2,+1}
    \end{pmatrix},\quad
    \Psi^{+}\equiv
    \begin{pmatrix}
        \chi_{1,0}\\
        \overline{\chi}_{2,0}
    \end{pmatrix},\quad
    \Psi^{++}\equiv
    \begin{pmatrix}
        \chi_{1,+1}\\
        \overline{\chi}_{2,-1}
    \end{pmatrix},
\eeq
the charge-conjugate fields are identified with $\Psi^{-}$ and $\Psi^{--}$. Observe the doubly charged states in the spectrum, in contrast to the $Y=0$ case.

The diagonal $Z$ couplings are
\beq
    g^{\Psi,Z}_{V,Q} \Big|_{\shortstack{\footnotesize $n=3$\\ \footnotesize $Y=1$}}=\frac{e}{s_W c_W}\Big(\mathcal{T}-(\mathcal{T}+1)s_W^2\Big).
\eeq
In particular, the neutral component (${\mathcal{T}}=-1$, $Q=0$) has a non-zero coupling to the $Z$ given by
\beq
    {g^{\Psi,Z}_{V,0}}    \Big|_{\shortstack{\footnotesize $n=3$\\ \footnotesize $Y=1$}}=- \frac{e}{s_W c_W}.
\eeq
For charged states, the couplings are
\beq
    {g^{\Psi,Z}_{V,\pm1}}\Big|_{\shortstack{\footnotesize $n=3$\\ \footnotesize $Y=1$}} &=\pm\frac{es_W}{c_W},\\
   {g^{\Psi,Z}_{V,\pm2}}\Big|_{\shortstack{\footnotesize $n=3$\\ \footnotesize $Y=1$}} &=\pm\frac{e}{s_W c_W}\Big(1-2s_W^2\Big).
\eeq
Similarly to the doublet, taking $Y=-1$ instead simply relabels which $\mathcal{T}$ component is electrically neutral.

\newpage
\subsection{Quintuplet Representation}
Moving to higher representations, the quintuplet corresponds to the $n=5$ representation, with isospin $T=2$ thus $\mathcal{T}\in\{-2,-1,0,+1,+2\}$. We first consider the case $Y=0$ and thus $Q=\mathcal{T}$. Accordingly, the Dirac fermions are
\beq\notag
    \Psi^{0}\equiv
    \begin{pmatrix}
        \chi_{1,0}\\
        \overline{\chi}_{2,0}
    \end{pmatrix},\quad
    \Psi^{+}\equiv
    \begin{pmatrix}
        \chi_{1,+1}\\
        \overline{\chi}_{2,-1}
    \end{pmatrix},\quad
    \Psi^{++}\equiv
    \begin{pmatrix}
        \chi_{1,+2}\\
        \overline{\chi}_{2,-2}
    \end{pmatrix},
\eeq
with the remaining components forming $\Psi^{-}$ and $\Psi^{--}$. For $Y=0$, it follows from Eq.~\eqref{eq:couplings} that the diagonal $Z$ couplings are identical to Eq.~\eqref{15}. Similarly, for the charged states 
\beq
    {g^{\Psi,Z}_{V,\pm1}}=\pm \frac{e c_W}{s_W},\qquad {g^{\Psi,Z}_{V,\pm2}}&=(\pm2) \frac{e c_W}{s_W}.
\eeq

The quintuplet is highlighted as a metastable representation in the original `Minimal Dark Matter' \cite{Cirelli:2005uq}. However, since the decay rate of particles typically scales with the mass of the decaying state, the strong stability statements which apply for the orthodox TeV-scenario must be re-examined. As we show below the $\boldsymbol{5}$ representations are not found to be automatically stable for the entire parameter space for which they can account for the dark matter relic abundance. 

For the quintuplet model ($n=5$) with $Y=0$, the $\Psi^0$ states that constitute the dark matter can decay via dimension-six operators of the form 
\beq
    \mathcal O_{n=5}^{Y=0} &\sim \frac{1}{\Lambda^2} \chi_{1} (LHH\tilde H),
\eeq
where $\tilde H=i\sigma_2 H^*$ and $\Lambda$ is the scale at which these operators are generated. Observe that the two fields $\chi_{1,0}$ and $\overline{\chi}_{2,0}$, which comprise $\Psi^0$, involve slightly different contractions with the Standard Model fields to form gauge invariant operators, but after EWSB these subtleties become irrelevant.\footnote{Mass shifts to  $\chi_1^0$ and $\chi_2^0$ from these operators are negligible, since  $\Delta\equiv m_1-m_2$ comes only via mixing with the Standard Model leptons such that $\Delta \sim v^6/(\Lambda^4 m )$. For high scale $\Lambda$ (and we anticipate $\Lambda\sim \Mpl$) this will {\em not} lead to the `pseudo-Dirac' direct detection suppression discussed in Footnote~\ref{fn1}.} After EWSB, the neutral-component couplings take the form
\beq
\mathcal L_{\rm eff}\supset
\frac{v^2}{\Lambda^2}\left(
 \chi_{1}^0 \nu h
+  \overline{\chi}_{2}^0 \nu h
\right)+\cdots ,
\eeq
permitting dark matter decays via $\chi^0\rightarrow h\nu$. There is also a mixing-induced coupling of $\chi$ to the electroweak gauge bosons (which can be seen by calculating the mass eigenstates of the $\chi$-$\nu$ system), which leads to (similarly suppressed) decay channels through $\nu Z$ and $l W$. 

For $m  \gg m_W,m_Z,m_h$, the total decay width is parametrically
\beq
    \Gamma_{\chi^0} \sim \frac{3}{8\pi} \frac{v^4}{\Lambda^4} m\,,
\eeq
where we include a factor of $3$ to account for the multiplicity of decay channels. Accordingly, the dark matter lifetime for $n=5$ and $Y=0$ is $\tau =1/\Gamma_{\chi^0}$. The characteristic limit on dark matter decaying in the manner above is $\tau \gtrsim  10^{30}$~s \cite{Das:2023wtk, Deligny:2024fyx}, which implies
\beq \label{cID1}
    m_{5}^{(Y=0)}\lesssim 5\times10^{10}~{\rm GeV}\left(\frac{\Lambda}{\Mpl}\right)^4.
\eeq
We return to calculate this limit with care in Section~\ref{sec3}.

In the above we took $n=5$ and $Y=0$, had we taken $Y=\pm1,+2$ then this also leads to decay operators arising at mass dimension six. For $Y=1$ the leading operator is $\chi L(H\tilde H\tilde H)$ and  for $Y=2$ it is $\chi (L\tilde H\tilde H\tilde H)$. However, for $Y=-2$ (the only remaining choice consistent with a dark matter candidate) the leading decay operator arises at mass dimension seven, namely
\beq
\mathcal O_{n=5}^{Y=-2} \sim \frac{1}{\Lambda^3}\,\chi\,e^c(H H H \tilde H).
\label{y=-2}
\eeq
The corresponding lifetime is  $\tau_{5}^{(Y=-2)} \propto(1/m)(\Lambda/v)^6$. Taking the characteristic limit $\tau \gtrsim  10^{30}$~s \cite{Das:2023wtk, Deligny:2024fyx} this lifetime limit implies a mass bound 
\beq\label{cID2}
    m_{5}^{(Y=-2)}\lesssim 5\times10^{42}~{\rm GeV}\left(\frac{\Lambda}{\Mpl}\right)^6.
\eeq
For $\Lambda\sim \Mpl$ decaying dark matter limits do not lead to a meaningful mass bound in this case. We can rephrase the question to ask how low the cut-off can be before the dark matter mass is constrained. Reinterpreting Eq.~\eqref{cID2}, we note that even for the heaviest dark matter $m\sim \Mpl$ for $\Lambda\gtrsim 2\times 10^{14}$~GeV the mass is unconstrained by searches for decaying dark matter for $n=5$ and $Y=-2$. 

\subsection{Septuplet Representation}
The final example we consider is the septuplet representation, which corresponds to $n=7$, thus $T=3$ and $\mathcal{T}\in\{-3,-2,-1,0,+1,+2,+3\}$. We will consider the $Y=0$ case for which $Q=\mathcal{T}$. The corresponding Dirac fermions are: $\Psi^{0},\Psi^{\pm},\Psi^{\pm\pm},\Psi^{\pm\pm\pm}$.
The diagonal $Z$ couplings follow from Eq.~\eqref{eq:couplings}
\beq
    g^{\Psi,Z}_{V,Q}=\frac{e c_W}{s_W} \mathcal{T},\qquad\qquad g^{\Psi,Z}_{A,Q}=0,
\eeq
again, we have $g^{\Psi,Z}_{V,0}=0$ since we consider the case $Y=0$. 

This representation can be sufficiently metastable to be a viable dark matter candidate without any additional stabilizing symmetry. If  $\Psi^0$ decays are induced from an intermediate scale $\Lambda$ this will generically lead to  dimension eight operators of the form
\beq
    \mathcal O_{n=7}^{Y=0}&\sim \frac{1}{\Lambda^4} \chi_1 L H H H \tilde H \tilde H,
\eeq
and after EWSB this induces $1\rightarrow2$ decay operators for $\Psi^0$ similar to the $\boldsymbol{5}$ case, with $\tau \sim \frac{8\pi}{3} (\Lambda^8/v^8 m)$. This implies the following restriction on the pair $m$, $\Lambda$
\beq\label{cID4}
    m\lesssim 10^{18}~{\rm GeV}\left(\frac{\Lambda}{2\times10^{11}~{\rm GeV}}\right)^8.
\eeq 
We see that dark matter is unconstrained by searches for decaying dark matter provided that $\Lambda\gtrsim 2\times10^{11}$~GeV. This intermediate mass scale is reminiscent of scales invoked with the PQ mechanism \cite{Peccei:1977hh} and the neutrino seesaw scale \cite{Yanagida:1979as}, so it remains an interesting prospect that signals might arise in future indirect detection experiments from the septuplet model.  

\section{Phenomenology}
\label{sec3}
Having established the relevant details for our models of interest, we proceed to study the phenomenological implications. In this section we calculate the dark matter relic abundance in the instantaneous reheating approximation, along with the associated limits from direct and indirect detection experiments.

\subsection{Dark Matter Relic Abundance}
The masses of the $\Psi$ states are split via radiative corrections which lead to the charged states picking up a small positive correction, the size of which is dependent on the dimension of the representation and the hypercharge. Specifically, the mass splitting between $\Psi^\pm$ and $\Psi^0$ is given by \cite{Cirelli:2005uq}
\beq\notag
    \Delta\equiv m_\pm-m_0 \simeq
    \begin{cases}
        341~{\rm MeV}  &\qquad n=2,~Y=1/2,\\
        525~{\rm MeV}  &\qquad n=3,~Y=1,\\
        166~{\rm MeV}  &\qquad n=3,~Y=0,\\
        166~{\rm MeV}  &\qquad n=5,~Y=0,\\
        166~{\rm MeV}  &\qquad n=7,~Y=0.
    \end{cases}
\eeq
The splitting is so small that elsewhere we do not make a distinction between the mass of the components and simply use $m\equiv m_0\simeq m_Q$.

Importantly, the EM neutral state $\Psi^0$ is always the lightest of the $\Psi$ states. The $\Psi$ states carry a conserved quantum number, either by construction (i.e.~via a $\mathbb{Z}_2$) or by accident. Hence, the heavier $\Psi$ states will decay to the neutral states. For the singly charged $\Psi^\pm$ this occurs via $\Psi^\pm\rightarrow \pi^\pm\Psi^0$, while for states with multiple electric charges, the decay to $\Psi^0$ is through a ladder of  off-shell $W^\pm$ cascades. Consequently, the relic abundance of $\Psi^0$ dark matter $Y^0_{\infty}$ is the sum of the freeze-in abundances of all the $\Psi$ states:
\beq\label{eq:0}
    Y_{\infty}^0=\sum_QY_Q\Big|_{\rm FI}~.
\eeq
In particular, freeze-in production can proceed in two distinct manners for each state
\begin{itemize}
    \item Neutral channels $q\bar q\to (\gamma/Z)\to \Psi_{Q} \bar\Psi_{Q}$
    \item ``Co-production'' of adjacent isospin states via charged mediators e.g~$u\bar d\to W^+\to \Psi_{Q+1}\bar\Psi_{Q}$.
\end{itemize}
It can be seen from Eq.~(\ref{eq:couplings}) that the cross-section for the $\Psi$ production via $Z$ mediation scales as (neglecting $Y$)
\beq
    \sigma(q\bar q\to \Psi_{Q}\bar\Psi_{Q})\propto (g^{\Psi,Z}_{V,Q})^2 \propto \mathcal{T}^2.
\eeq
Thus, higher isospin components are produced more readily (and will subsequently decay to $\Psi^0$). It follows that the production rate of the $\mathcal{T}^{\rm th}$ pair with zero net-charge $\Psi^{Q}\bar\Psi^{Q}$ compared to the pair of singly charged states\footnote{Choosing the reference cross-section to correspond to $\Psi^0$ is dangerous since for $Y=0$ the tree-level cross-section vanishes.} is parametrically
\beq\notag
\sigma_{q\bar q\to \Psi_{Q}\bar\Psi_{Q}}
\simeq \mathcal{T}^2\times \sigma_{q\bar q\to \Psi^-\Psi^+}.
\eeq
Similarly, the charged channel co-production $\Psi_{Q}\bar\Psi_{Q-1}$ (with isospins ${\mathcal{T}}$ and ${\mathcal{T}} - 1$)  scales as
\beq\notag
    \sigma_{q\bar q'\to \Psi_{Q}\bar\Psi_{Q-1}} \simeq (T-(\mathcal{T}-1))(T+\mathcal{T}) \sigma_{q\bar q'\to \Psi^+\bar\Psi^0},
\eeq
summing over all quarks and taking the appropriate sign final state to match the initial state electric charge. To a good approximation, we can just include intra-generation $W$ processes, as alternative charged channels are CKM suppressed. A more detailed treatment will lead to only an $\mathcal O(1)$ correction. In our numerical results, we include all tree-level cross-sections. 

To proceed, we should calculate the freeze-in abundances for a general $\Psi_{Q}$, and then sum these up to obtain the relic abundance, cf.~Eq.~\eqref{eq:0}. The relic abundance for  $\Psi_{Q}$ is obtained by evaluating the Boltzmann equation
\beq \label{eq:BE}
    \dot n_Q + 3  H  n_Q = \gamma_{Q}(T) .
\eeq
Neglecting quark masses, the reaction density for $q\bar q'\to \Psi_Q\bar\Psi_{Q'}$ is given by
\beq
\notag
  \gamma_{Q} = \frac{T}{32\pi^4} \sum_{qq'}\int_{4m^2}^{\infty} {\rm d}s \Big[ (s-4m^2)\sqrt{s} \sigma_{q\bar q'\to {\Psi_Q}{\bar\Psi_{Q'}}}  \Big] K_1 \left(\frac{\sqrt{s}}{T}\right),
\eeq
in terms of the modified Bessel function  $K_1$.

Since we are working in the  Boltzmann suppressed freeze-in regime $T\ll m$, production is dominated near-threshold. Thus, we parameterize the center-of-mass energy as $s\simeq 4m^2+\varepsilon$ taking $\varepsilon\ll 4m^2$ and hence
\beq
    \beta\simeq\sqrt{\frac{\varepsilon}{4m^2}}\ll1.
\eeq
Moreover, $(s-4m^2)\, \sqrt{s}\, \sigma(s)$ as appears in the expression for $\gamma_{Q}$ (in the square brackets) can be approximated by
\beq
    (s - 4  m^2)\, \sqrt{s}\, \sigma(s) \simeq  \left(\frac{\mathcal{K}^{q\bar q'}_{QQ'}}{96\pi m^{2}} \right) \varepsilon^{3/2},
\eeq
where  we recall $\mathcal{K}^{q\bar q'}_{QQ'}$ collects all the various couplings together, cf.~Eq.~\eqref{10}. Taking the Bessel function large value approximation $K_1(z) \simeq \sqrt{\frac{\pi}{2z}}  e^{-z}$ for $z\gg1$, and working to leading order in $\varepsilon$ yields
\beq
    K_1\left(\frac{\sqrt{s}}{T}\right) \simeq \sqrt{\frac{T\pi}{4m}} e^{-\frac{2m}{T}} e^{-\frac{\varepsilon}{4m T}}~.
\eeq
It follows that the reaction density can be rewritten as an integral over $\varepsilon$ in the following manner
\beq\notag
    \gamma_{Q} \simeq  \sum_{q,q'}\frac{T}{32\pi^4}  \frac{\mathcal{K}^{q\bar q'}_{QQ'}}{96\pi m^{2}} \sqrt{\frac{\pi T}{4m}} e^{-\frac{2m}{T}} \int_0^\infty {\rm d}\varepsilon ~ \varepsilon^{3/2} e^{-\frac{\varepsilon}{4m T}}.
\eeq
Evaluating the $\varepsilon$ integral yields the reaction for $m \gg T$
\beq\label{111}
    \gamma_{Q} \simeq \sum_{q,q'}\left(\frac{\mathcal{K}^{q\bar q'}_{QQ'}}{256 \pi^4} \right)T^{4} e^{-\frac{2m}{T}},
\eeq
where we must evaluate the combined coupling $\mathcal{K}^{q\bar q'}_{QQ'}$. 

Returning to the Boltzmann equation, Eq.~\eqref{eq:BE}, we can use the form of $\gamma_Q(T)$ to calculate the freeze-in yields $Y \equiv n_Q/s$ for each component via 
\beq \label{eq:yield}
    \frac{dY}{dT} =  - \frac{\gamma_{Q}(T)}{s  H  T} ,
\eeq
where  $ H(T) = \sqrt{\frac{\pi^2 \gs}{90}}   \frac{T^2}{\Mpl}$ and $s(T) = \frac{2\pi^2}{45}\gss T^3$ in terms of the effective entropy degrees of freedom $\gss$, and the relativistic degrees of freedom $\gs$, and the reduced Planck mass $\Mpl \simeq 2.4 \times 10^{18}$~GeV. Substituting $\gamma_Q(T)$ from Eq.~\eqref{111}, it follows that
\beq\notag
    \frac{dY}{dT} = - \frac{135\sqrt{10}}{512 \pi^7} \sum_{q,q'}\left( \frac{\mathcal{K}^{q\bar q'}_{QQ'}}{\gss\sqrt{\gs}}  \right)\frac{\Mpl}{T^2}  e^{-\frac{2m}{T}}.
\eeq

We assume that the initial abundances  of $\Psi$ states are negligible and that there is no appreciable $\Psi$ production from direct inflaton decays. We work in the Boltzmann suppressed regime $m\gg \Trh$ and restrict our attention (in this section) to the case of {\it instantaneous} reheating after inflation to a reheat temperature $\Trh$. In this setting, the freeze-in yield of the $\Psi$ state with charge $Q$ is given by
\beq
    Y_{\rm FI}^\text{inst} &= \frac{135\sqrt{10}}{512 \pi^7} \sum_{q,q'}\frac{\mathcal{K}^{q\bar q'}_{QQ'}}{\gss\sqrt{\gs}} \Mpl \int_0^{\Trh} dT  \frac{e^{\frac{-2m}{T}}}{T^2}.
\eeq
Integrating gives the yield (for the case of instantaneous reheating)
\beq
    Y_{\rm FI}^\text{inst} &\simeq \frac{135\sqrt{10}}{1024 \pi^7} \sum_{q,q'} \frac{\mathcal{K}^{q\bar q'}_{QQ'}}{\gss\sqrt{\gs}} \frac{\Mpl}{m} e^{-\frac{2m}{\Trh}}.
    \label{eq:Y0}
\eeq

The freeze-in yields scale as $Y_{\rm FI} \propto \exp(-2m/\Trh)$, as anticipated for  Boltzmann suppressed freeze-in ($m\gg \Trh$). From $Y_{\rm FI}$ we can calculate the freeze-in yield of each Dirac fermion by taking the appropriate $ \mathcal K$ in turn, as outlined in Eq.~\eqref{9}. We recall that the numerical values for $ \mathcal K$ are provided in Tables~\ref{tab:CZ}-\ref{tab:CW} for selected models of interest. 
The yield can be subsequently compared with the observed relic abundance by noting that $\Omega_{\rm DM}= m\, Y_{\rm FI}\, s_0/\rho_c$ where $s_0$ is the entropy density today and $\rho_c$ is the critical density.

In the instantaneous reheating approximation, for a fixed representation, the model has only two degrees of freedom $\Trh$ and $m$. As a result, this class of models is highly predictive. Note that the exponential in Eq.~\eqref{eq:Y0} strongly determines the value of $Y_{\rm FI}^{\rm inst}$, thus it is the ratio $m/\Trh$ that is critical to reproducing the observed relic abundance. In Table~\ref{tab:4} we provide the approximate value which gives the correct value of $\Omega_{\rm DM}$ in each of our models of interest. Observe that for low $n$ electroweak representations one typically requires $m/\Trh \simeq 23 \pm 1$. 
\begin{table}[t]
\centering
    \renewcommand{\arraystretch}{0.95}
    \begin{tabular}{lc}
        \hline\hline
            & $m/\Trh$ \\
        \hline
        Doublet            & 22.4 \\
        Triplet $(Y=1)$    & 23.2 \\
        Triplet $(Y=0)$    & 23.0 \\
        Quintuplet $(Y=0)$ & 23.8 \\
        Septuplet $(Y=0)$  & 24.3 \\
        \hline\hline
    \end{tabular}
    \caption{Values of $m/\Trh$ required to fit the whole observed dark matter abundance (assuming instantaneous reheating), for the different SU(2)$_L$ representations. \label{tab:4}}
\end{table}

\begin{figure}[t!]
    \def\sepf{0.95}
    \centering
    \includegraphics[width=\sepf\columnwidth]{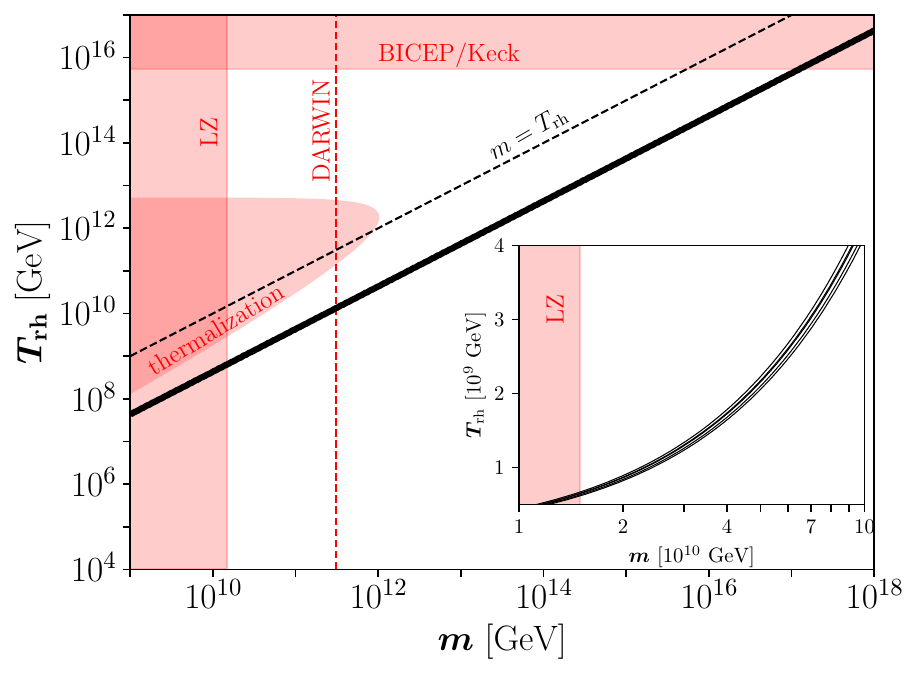}
    \vspace{-6mm}
    \caption{The solid lines correspond to the reheat temperature $\Trh$ which gives the observed dark matter relic density of mass $m$ assuming instantaneous reheating, for the different SU(2)$_L$ representations. The inset plot shows a zoom of the different lines. The vertical red regions indicate the parameter space excluded by direct detection (LZ); the projected sensitivity of DARWIN is also shown. The black dashed line indicates $\Trh = m$, above which freeze-in is no longer Boltzmann suppressed. The red shaded ``thermalization'' region indicates parameter values for which $\Psi^0$ would enter equilibrium with the Standard Model.}    \vspace{-3mm}
    \label{fig:instantaneous}
\end{figure}

In Figure~\ref{fig:instantaneous} we show the reheat temperature $\Trh$ required to reproduce the observed dark matter relic abundance as a function of the dark matter mass $m$. In calculating the relic abundance, we sum over all the quark initial states (neglecting other channels) and all the $\Psi$ final states and then account for charged $\Psi$ decays to the (meta)stable neutral state $\Psi^0$. Observe in Figure~\ref{fig:instantaneous} that all of the low $n$ models studied here have comparable relic density curves, indeed they are only distinguishable in the `zoomed' subpanel. In Section~\ref{sec5} we study the impact of deviating from instantaneous reheating. 

\subsection{Cosmological Bounds}
\vspace{-2mm}
Having demonstrated that the relic abundance $\Omega_{\rm DM}\, h^2 \simeq 0.12$ can be reproduced, we turn to the limits on these models. We first examine the observational and consistency constraints that arise from cosmology. The reheat temperature is constrained at the higher and lower ranges. The cosmic abundances of nuclei agree well with the predictions of Big Bang Nucleosynthesis (BBN). Consistency with BBN requires that the Standard Model thermal bath was in thermal equilibrium at temperatures around a few MeV. This implies a lower bound on the reheat temperature \cite{Sarkar:1995dd}:
\beq
    \Trh \gtrsim 5~{\rm MeV}.
\eeq

The upper temperature limit of $\Trh$ is constrained by cosmological observations by BICEP/Keck \cite{BICEP:2021xfz}. Specifically, BICEP/Keck searches for primordial B-modes constrain the tensor-to-scalar ratio $r$. This can be related to the Hubble scale during inflation $H_I$ \cite{Baumann:2009ds}
\beq
    H_I  =  \pi\, \Mpl\sqrt{r\, A_s/2},
\eeq
where $A_s\simeq 2.1\times 10^{-9}$ is the measured scalar amplitude. Applying the BICEP/Keck bound $r<0.036$ \cite{BICEP:2021xfz} one obtains an upper bound on $H_I$ \cite{Baumann:2009ds}
\beq \label{eq:HI}
    H_I  \lesssim  6\times 10^{13}~{\rm GeV}.
\eeq
This in turn constrains radiation energy density and hence the maximum temperature of the thermal bath via
\beq
    \rho_{\rm rh} = \frac{\gs(\Trh) \pi^2}{30}\, \Trh^4 = 3\Mpl^2 H_{\rm rh}^2 \le 3\Mpl^2 H_I^2.
\eeq
Assuming instantaneous reheating, it follows that
\beq\notag
    \Trh \lesssim 6\times10^{15}~{\rm GeV} \left(\frac{106.75}{\gs}\right)^{1/4} \left(\frac{H_I}{6\times10^{13}~{\rm GeV}}\right)^{1/2},
\eeq
 corresponding to the upper red band in Figure~\ref{fig:instantaneous}.

Finally, we should identify the parameter range for which the dark matter enters equilibrium with the thermal bath of Standard Model particles. Equilibration is avoided, provided that the production rate is smaller than the expansion rate $H$, this restriction can be expressed as
\beq
    \gamma(\Trh) < n_\text{eq}(\Trh) H(\Trh),
\eeq
where $n_{\rm eq}$ is the equilibrium number density of dark matter. The region in which thermalization occurs is marked in Figure~\ref{fig:instantaneous}. Note that the line $m = \Trh$ separates the relativistic and non-relativistic regimes. The inflection in the bounding curve of the thermalization region corresponds to the transition from relativistic to non-relativistic regimes.  Furthermore, we highlight that these cosmological limits do not meaningfully constrain the parameter space of our models of interest. 

\vspace{-2mm}
\subsection{Indirect detection}
\vspace{-2mm}
Experimental observations of extragalactic $\gamma$-rays and cosmic rays constrain dark matter decays and annihilation. 
In models with $Y=0$ TeV scale masses are allowed without conflict with direct detection, in this case searches for dark matter annihilation place lower limits on the mass.
The analysis of indirect detection signals for annihilating dark matter arising from electroweak dark matter must take into account Sommerfeld enhancement of the annihilation cross-section \cite{Hisano:2004ds, Feng:2010zp}. An updated analysis is beyond the scope of this work, so we draw on existing analyses
\beq \label{eq:mID}
    m \gtrsim
    \begin{cases}
        3.5~{\rm TeV} &~~ n=3 ~~\text{\cite{Safdi:2025sfs}},\\
        20~{\rm  TeV} &~~ n=5 ~~\text{\cite{Garcia-Cely:2015dda}},\\
        20~{\rm  TeV} &~~ n=7 ~~\text{\cite{Garcia-Cely:2015dda, Panci:2024oqc}}.
    \end{cases}
\eeq
We highlight that these limits derived in Ref.~\cite{Garcia-Cely:2015dda} are based on older HESS data \cite{HESS:2013rld}, while recent analyzes have found improvements using Fermi data \cite{Safdi:2025sfs, Aghaie:2025iyn} (although restricted to the $\sim$TeV mass range). Moreover, the exact numbers are dependent on the assumed dark matter halo profile (the above assume the Einasto profile \cite{Einasto:1965czb}).

\begin{figure}[t!]
    \def\sepf{0.9}
    \centering
    \includegraphics[width=\sepf\columnwidth]{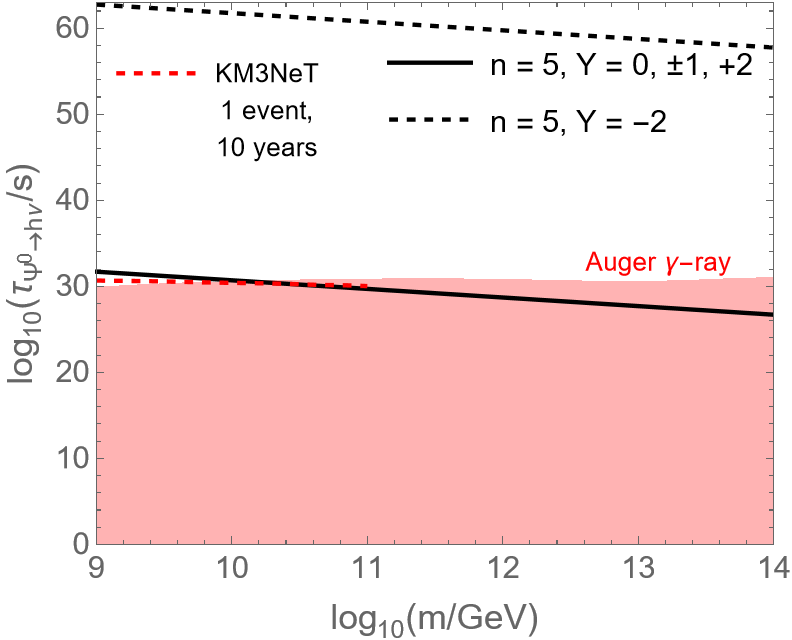}
    \caption{We show the lifetimes of the quintuplet ($\boldsymbol{5}$) representations due to Planck induced higher dimension operators for the two distinct cases $Y=0,\pm1,+2$ and $Y=-2$. This is plotted as a function of mass $m$. We overlay the indirect detection limits for decaying dark matter coming from Auger's observation of galactic $\gamma$-rays in red, specialized for the decay channel $\Psi^0 \to h \nu$ \cite{Deligny:2024fyx}. For $Y=0,\pm1,+2$ Auger places an upper bound on the dark matter mass. We also show the anticipated reach of KM3NeT (based on the assumption of a single event
    over 10 years) \cite{Kohri:2025bsn} as the red dotted curve.}
    \label{fig:ID}
\end{figure}

Furthermore, while the $\boldsymbol{5}$ and $\boldsymbol{7}$ representations can be naturally metastable without an additional (e.g.~$\mathbb{Z}_2$) stabilizing symmetry, in the high mass limit, decays of $\Psi^0$ descending from $\boldsymbol{5}$ are constrained by decaying dark matter bounds for $Y=0,\pm1,+2$ (unless an additional stabilizing symmetry is assumed).
Recall from Section~\ref{sec4}, that the $\Psi^0$ decays dominantly to $h\nu$. In this case Auger places a lower bound on the dark matter lifetime of order $10^{30}$ s. We provided  analytic estimates of the lifetime constraints in Eqs.~\eqref{cID1}, \eqref{cID2} and \eqref{cID4}, in Figure~\ref{fig:ID} we present the detailed bounds, drawing on the channel specific analysis in \cite{Deligny:2024fyx}.
These limits are comparable to the bounds on other decay channels presented in Ref.~\cite{Das:2023wtk}.

For $Y=0, \pm1,2$, decays of $\Psi^0$ induced by Planck-suppressed operators (saturating $\Lambda \sim \Mpl$ in Eq.~\eqref{cID1}) are constrained by Pierre Auger observations of galactic $\gamma$-rays, yielding an upper mass bound
\beq
    m \lesssim 1.2 \times 10^{10}~{\rm GeV}.
\eeq
This is in agreement with our earlier analytic estimates. We note that for $Y = - 2$ the dark matter $\Psi^0$ is cosmologically stable (and thus unbounded) unless the decay operators are induced well below the Planck scale (i.e.~$\Lambda\ll \Mpl$), as  described by Eq.~\eqref{y=-2}.

Finally, in Figure \ref{fig:ID} we also indicate the expected reach of next-generation indirect detection experiments to constrain the dark matter lifetime, specifically KM3NeT  (1 event, 10 years) \cite{Kohri:2025bsn}, although this seemingly offers limited improvement.  Moreover, it is anticipated that the Cherenkov Telescope Array Observatory (CTAO) will strengthen the annihilation bounds on dark matter coming from $\boldsymbol{5}$ ($\boldsymbol{7}$) representation, with a reach of around 30~TeV (75~TeV) \cite{Garcia-Cely:2015dda} (see also Ref.~\cite{Abe:2025lci}). Thus, one expects an improvement in the lower mass bounds on these models in the near-future.

\subsection{Direct Detection}
\vspace{-2mm}
Next, we examine the constraints that arise from direct detection, most prominently from the LZ experiment \cite{LZ:2024zvo}. Comparable constraints have also been derived by XENONnT  \cite{XENON:2025vwd} and PandaX  \cite{PandaX:2024qfu}; we also highlight the LZ dedicated high mass analysis \cite{LZ:2024psa}. To proceed we should calculate the direct detection cross-sections for the dark matter models arising in Section~\ref{sec4}. The cases with $Y\neq0$ lead to tree-level $Z$ couplings for the dark matter $\Psi^0$ in which case the cross-section for $\Psi^0$ to scatter on an individual nucleon $N = p$, $n$ is given by \cite{Essig:2007az}
\beq\label{eq:SI}
    \sigma_{\rm SI}^{(N)} = \frac{\mu_N^2}{\pi} \left(\frac{ g^{\Psi,Z}_{V,0}  g_V^{N,Z}}{m_Z^2}\right)^2.
\eeq
The coupling $g^{\Psi,Z}_{V,0}$ is given in  Eq.~(\ref{eq:couplings}) for a vector-like pair of general SU(2)${}_L$ representation. Provided that $g^{\Psi,Z}_{V,Q=0}\neq0$, to compute $\sigma_{\rm SI}^{(N)}$ all that we require is the effective nucleon couplings $g_V^{N,Z}$. To find   $g_V^{N,Z}$ we first ascertain the couplings of the Standard Model quarks which arise from Eq.~\eqref{qqq} (with $T_3^{R}=0$), namely for $q=u,d$
\beq
    \label{q}
    {g^{u,Z}_{V}} &= \frac{e}{2s_W c_W}\Big(\frac{1}{2} - \frac{4}{3}s_W^2 \Big), \\[4pt]
    {g^{d,Z}_{V}} &=\frac{e}{2s_W c_W}\Big(-\frac{1}{2} + \frac{2}{3}s_W^2 \Big),\\[4pt]
    {g^{u,Z}_{A}} &= -   {g^{d,Z}_{A}} = \frac{e}{4s_W c_W}.
\eeq
We will take $\sin^2\theta_W|_{\mu=m_Z}\approx 0.231$ which is the $Z$ pole value, however, this does not change significantly via running to  intermediate scales \cite{Bernal:2026clv}. 
The coupling to protons ($p$) and neutrons ($n$) can then be obtained via
\beq
    {g^{p,Z}_{V}} &= 2  {g^{u,Z}_{V}} + {g^{d,Z}_{V}}\\[6pt]
    {g^{n,Z}_{V}} &= {g^{u,Z}_{V}} + 2  {g^{d,Z}_{V}}.
\eeq

From this we can write the direct detection cross-section for all cases of interest with $Y\neq0$
\beq\label{xs}\notag
    \sigma_{\rm SI} \simeq
    \begin{dcases}
        4.7\times10^{-40}~{\rm cm}^2  &\qquad  n=2,~Y=\pm1/2,\\
        1.9\times10^{-39}~{\rm cm}^2  & \qquad n=3,~Y=\pm1,\\
        1.9\times10^{-39}~{\rm cm}^2  & \qquad n=5,~Y=\pm1,\\
        7.6\times10^{-39}~{\rm cm}^2  &\qquad  n=5,~Y=\pm2.
    \end{dcases}
\eeq
If $Y=0$ the scattering cross-section is significantly reduced since the tree-level $Z$ coupling vanishes, and the elastic SI cross-section arises at loop-level \cite{Hisano:2004pv, Cirelli:2005uq, Chen:2023bwg}. In the following, we report the cross-sections computed in Ref.~\cite{Chen:2023bwg} for the $Y=0$ models\footnote{We thank the authors of Ref.~\cite{Chen:2023bwg} for providing numerical values.}
\beq\label{xs2}\notag
    \sigma_{\rm SI} \simeq
    \begin{cases}
        [3.4, 13]\times 10^{-48}~{\rm cm}^2 &\qquad n=3,~Y=0,\\
        [3.1,12]\times 10^{-47}~{\rm cm}^2 &\qquad n=5,~Y=0,\\
        [1.2, 4.8]\times 10^{-46}~{\rm cm}^2 &\qquad n=7,~Y=0.
    \end{cases}
\eeq
In particular, for $Y=0$ the scattering cross-sections are reduced by a factor of $10^8$ relative to their $Y=1$ counterparts. We note that $n=5$, $Y=1$ is identical to the triplet $n=3$, $Y=1$ since their couplings $g_{V,0}^{\Psi,Z}$ are identical, cf.~Eq.~\eqref{eq:ZPsi0}. 

\begin{figure*}[t!]
    \def\sepf{1}
    \centerline{\includegraphics[width=\sepf\columnwidth]{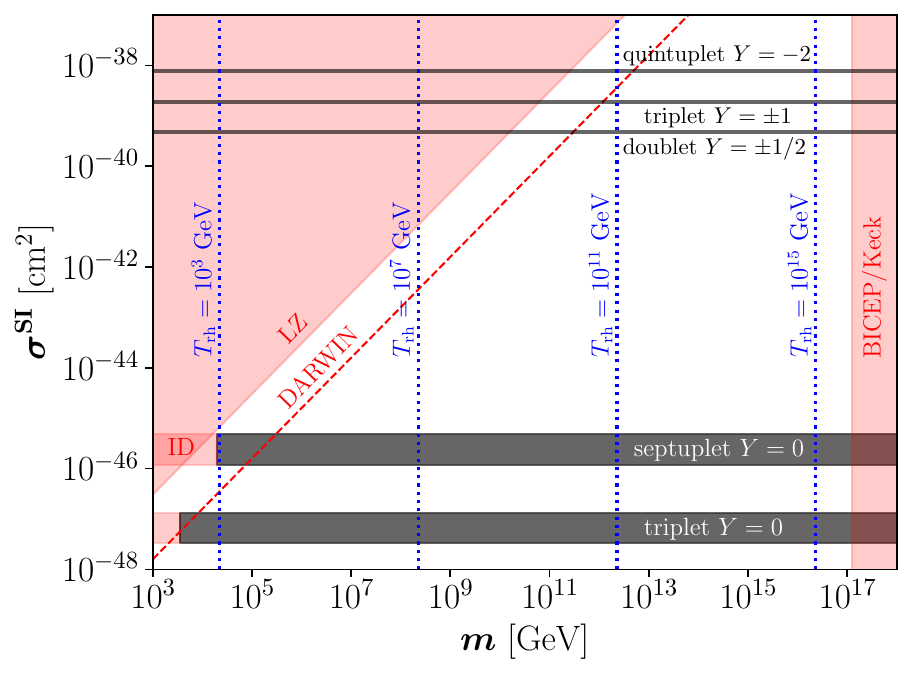}
    \includegraphics[width=\sepf\columnwidth]{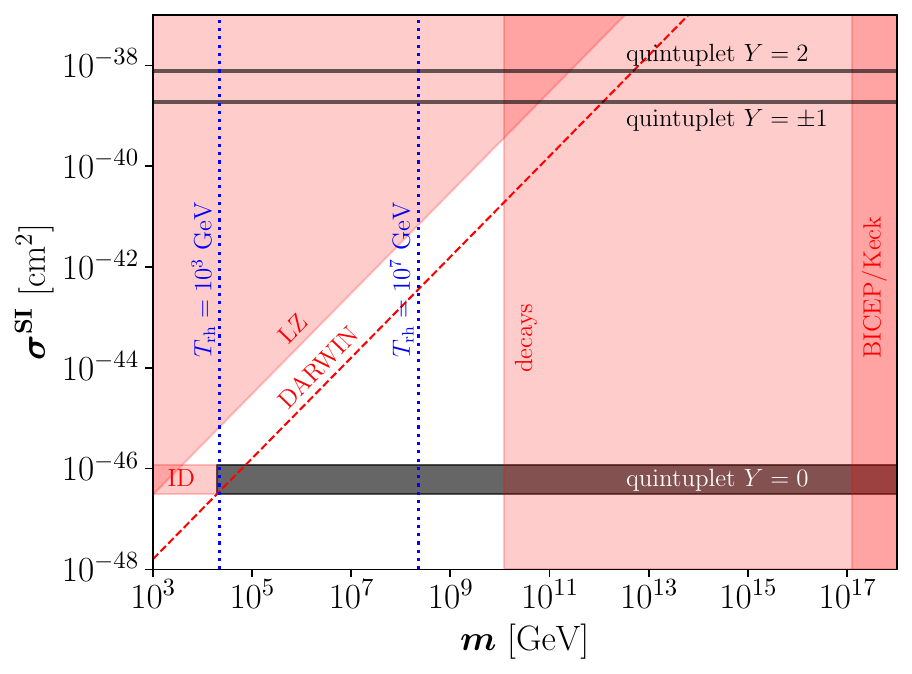}}
    \vspace{-2mm} 
    \caption{Constraints from spin-independent limits LZ, for the different representations. We also show the projections of the proposed DARWIN experiment. For instantaneous reheating the dark matter mass uniquely determines the required $\Trh$ and we plot some characteristic contours. We also collect together the cosmological limits from BICEP/Keck which constrain $\Trh$ and the bounds from dark matter annihilations (cf.~Eq.~(\ref{eq:mID})), marked `ID'. We separate the selection of models which are constrained by dark matter decays limits, which we show on the right hand panel, marked `decays' (cf.~Figure \ref{fig:ID}).}
    \label{fig:DD}
\end{figure*} 
Notably, the current constraint from LZ constrains the dark matter scattering cross-section to be \cite{LZ:2024zvo} 
\beq
    \left.\sigma_{\rm SI}^{(n,p)}\right|_{m\gg m_Z} < 3\times 10^{-47} {\rm cm}^2 \left(\frac{m}{{\rm TeV}}\right).
\eeq
We overlay this limit on Figure \ref{fig:instantaneous}. Moreover, in Figure~\ref{fig:DD} we show the constraints on the various models of interest in the familiar direct detection style plot, also combining the constraints from indirect detection and cosmology. In particular, smaller masses are constrained by dark matter annihilation in the cases of zero hypercharge for triplets, quintuplets and septuplets; cf. Eq. \eqref{eq:mID}. Additionally, an upper bound on the mass comes from dark matter decays in the case of the quintuplet (cf.~Figure~\ref{fig:ID}). The combination of direct detection and decaying dark matter bounds completely exclude the $n=5$ case with $Y=\pm1,2$. Since the $\boldsymbol{5}$ with $Y=-2$ is metastable for $\Mpl$ induced operators (cf.~Figure~\ref{fig:ID}) it remains a viable dark matter candidate provided $m \gtrsim 2 \times 10^{11}$ GeV.

We also mark the reach of DARWIN \cite{DARWIN:2016hyl} on Figures~\ref{fig:instantaneous} and \ref{fig:DD}. Notably, Darwin improves the bounds on dark matter by two orders of magnitude at all mass scales. The parameter space below DARWIN's reach is in the neutrino fog (cf.~Ref.~\cite{OHare:2020lva}) and novel experimental advances will be required for further progress.

\vspace{-2mm}
\section{Non-instantaneous Reheating}
\vspace{-2mm}
\label{sec5}
In our companion paper and the considerations above we have worked under the assumption that inflationary reheating is instantaneous. Notably, in this case the maximum temperature of the Universe is exactly the reheat temperature
\beq
    \Tmax\big|_{\rm instant} =  \Trh \simeq \sqrt{\Gamma_\phi \Mpl} ,
\eeq
where $\Gamma_\phi$ is the decay rate of the inflaton. However, beyond the instantaneous inflaton decay approximation, the maximum temperature of the thermal bath can exceed the reheat temperature \cite{Giudice:2000ex}, i.e.~$\Tmax> \Trh$. Moreover, in the regime $\Tmax> T > \Trh$ the freeze-in dynamics is sensitive to the equation-of-state parameter of the Universe. Indeed, details on how the energy density of the Universe is transmitted from the inflaton to the Standard Model plasma can radically alter expectations for freeze-in \cite{Garcia:2017tuj, Bernal:2019mhf, Garcia:2020eof, Bernal:2020bfj, Bernal:2020qyu, Allahverdi:2020bys, Garcia:2021iag, Barman:2022tzk, Batell:2024dsi, Bernal:2025fdr}. 

Suppose that during cosmic reheating, the dominant component of the Universe in terms of energy density has an equation-of-state parameter $\omega$, and that the Standard Model temperature scales as a power law. Let $\Trh$ and $\arh$ denote the Standard Model temperature and scale factor at the onset of the radiation-dominated era, respectively. Under these assumptions, the evolution of the Hubble expansion rate $H$ as a function of the scale factor $a$ is given by \cite{Bernal:2024yhu} 
\beq
    H(a) \simeq \Hrh \times
    \begin{dcases}
        \left(\frac{\arh}{a}\right)^\frac{3(1+\omega)}{2} &\text{ for } a \leq \arh ,\\
        \left(\frac{\arh}{a}\right)^2 &\text{ for } \arh \leq a ,
    \end{dcases}
\eeq
while the Standard Model bath temperature evolves as
\beq
    T(a) \simeq \Trh \times
    \begin{dcases}
        \left(\frac{\arh}{a}\right)^\alpha &\text{ for } a \leq \arh ,\\
        \left(\frac{\arh}{a}\right) &\text{ for } \arh \leq a .
    \end{dcases}
\eeq
In conventional cosmology $\alpha > 0$ so that the Standard Model temperature always decreases with time, as is typical (but in certain scenarios $\alpha \leq 0$ is possible \cite{Co:2020xaf}).

\begin{figure*}[ht!]
    \def\sepf{0.93}
    \centering
    \includegraphics[width=\sepf\columnwidth]{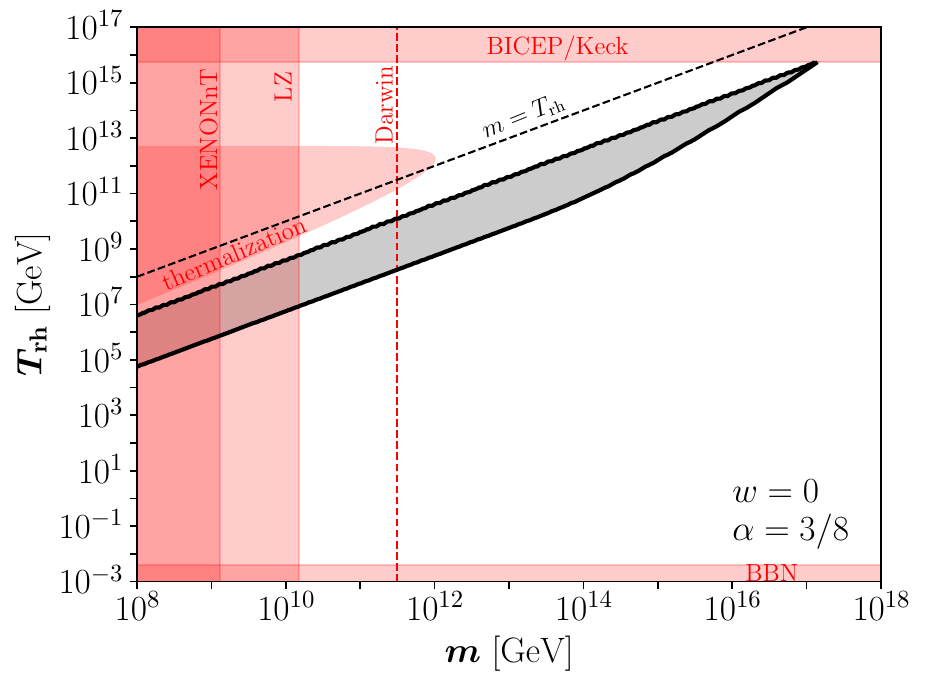}
    \includegraphics[width=\sepf\columnwidth]{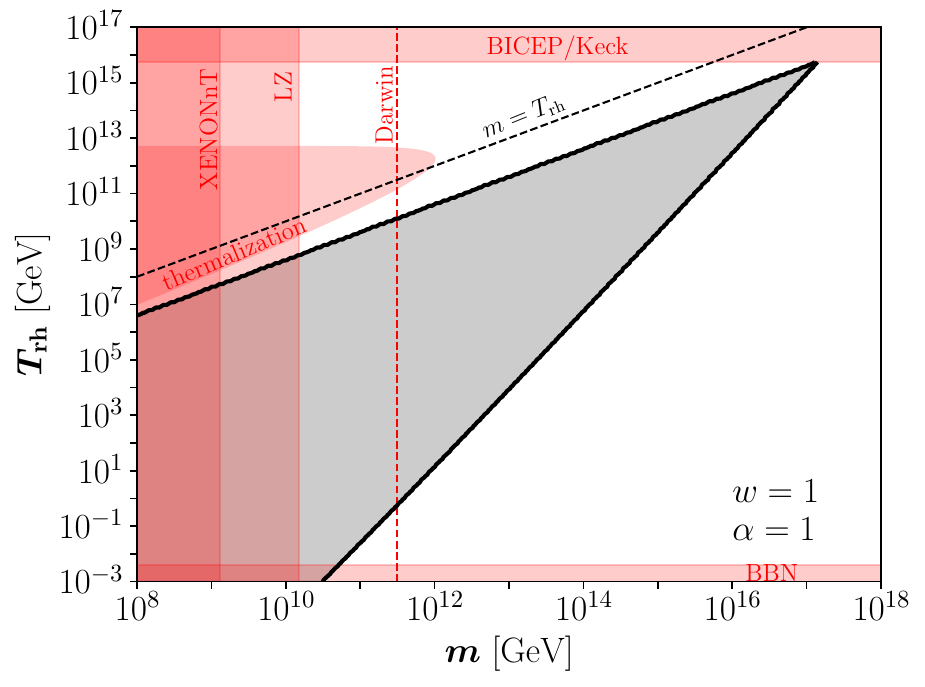}\\
    \includegraphics[width=\sepf\columnwidth]{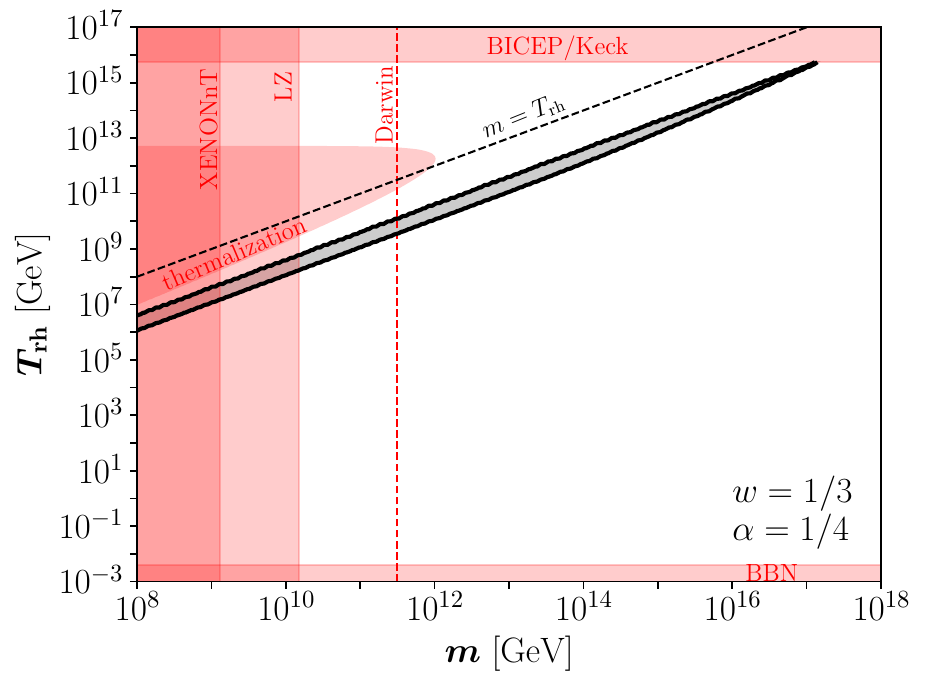}
    \includegraphics[width=\sepf\columnwidth]{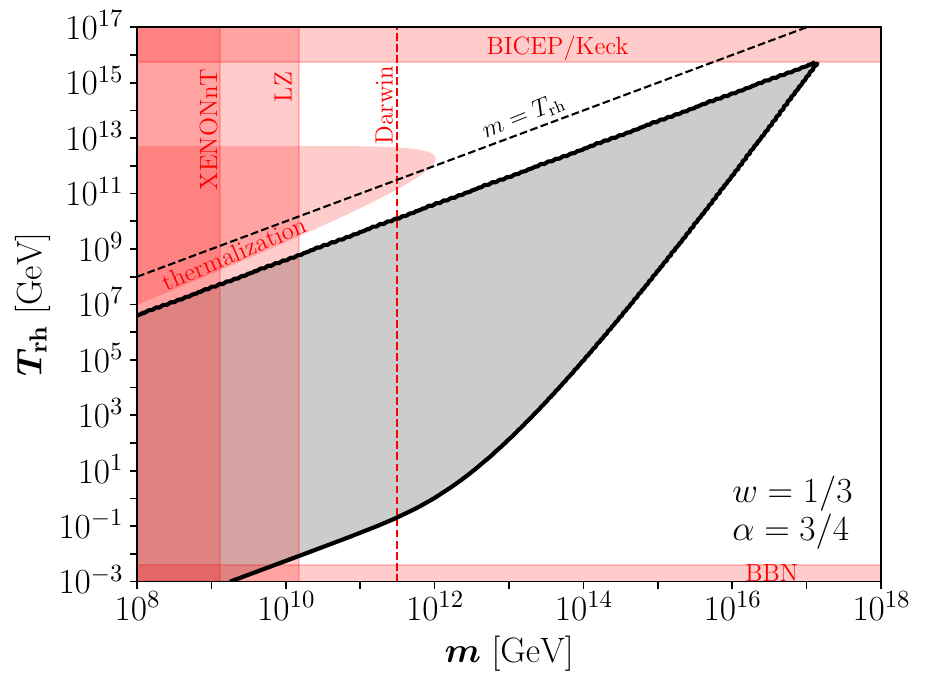}
    \vspace{-3mm}
    \caption{Non-instantaneous reheating.  The shaded region indicates viable parameters.  The upper boundary is set by instantaneous reheating ($T_I = \Trh$); the lower boundary corresponds to the maximal duration of reheating, specifically $H(T_I)< 6\times 10^{13}~{\rm GeV}$ consistent with BICEP/Keck \cite{BICEP:2021xfz} .The `thermalization' region indicates parameters for which the dark matter enters equilibrium with the thermal bath, in which case the observed relic abundance is not reproduced.}
    \label{fig:non-instantaneous}
\end{figure*} 

Taking into account that at the end of reheating $\rho_R(\arh) = \rho_\phi(\arh)$, the maximum energy density reached by the Standard Model thermal bath at the beginning of reheating (that is, at $a=a_I$) can be estimated and corresponds to a temperature $T_I\geq \Trh$ given by \cite{Barman:2021ugy, Bernal:2024yhu}
\beq \label{eq:Tmax}
    T_I \simeq \Trh \left[\frac{90}{\pi^2  \gs}  \frac{H_I^2   \Mpl^2}{\Trh^4}\right]^\frac{\alpha}{3 (1+\omega)}.
\eeq

In a general cosmic background where the Standard Model entropy is not conserved, instead of the yield $Y$, it is convenient to track the evolution of $N_Q \equiv n_Q\, a^3$. Therefore, for each component, the Boltzmann equation, Eq.~\eqref{eq:BE} can be expressed as
\beq \label{eq:dNda}
    \frac{dN_Q}{da} = \frac{a^2  \gamma_Q}{H} .
\eeq
The dark matter abundance produced during reheating (between $a_I$ and $\arh$), in the case where $\alpha \ne 0$, can be written as
\beq \label{eq:Yrh}
    Y_{\rm FI}^\text{rh} &\simeq \frac{405}{256  \pi^7}  \frac{\mathcal K^{q\bar q'}_{QQ'}}{  \gss  \alpha}  \sqrt{\frac{10}{\gs}}  \frac{\Mpl}{\Trh}\left(\frac{\Trh}{2  m}\right)^\eta \\
    &\qquad \times  \left[\Gamma\left(\eta, \frac{2 m}{T_I}\right) - \Gamma\left(\eta, \frac{2 m}{\Trh}\right)\right],
    \eeq
where $\Gamma(\eta,x)$ is the incomplete gamma function and
\beq
    \eta \equiv \frac{9+3\omega-8\alpha}{2\alpha}
\eeq
collects the cosmological parameters $\omega$ and $\alpha$.

In particular, when $\alpha > 0$ (corresponding to the thermal bath cooling with time, as typically occurs),  then Eq.~\eqref{eq:Yrh} can be approximated by
\beq \label{eq:YFI}
    Y_{\rm FI}^\text{rh} \simeq Y_{\rm FI}^\text{inst} \times
    \begin{dcases}
        \frac{1}{\alpha} \left(\frac{\Trh}{T_I}\right)^{\eta-1} e^{\frac{2 m}{\Trh} - \frac{2 m}{T_I}} & \\
        &\hspace{-1cm} \text{for } \Trh < T_I \ll m,\\
        \frac{\Gamma(\eta)}{\alpha} \left(\frac{\Trh}{2  m}\right)^{\eta-1} e^{\frac{2 m}{\Trh}} & \\
        &\hspace{-1cm}\text{for } \Trh < m \ll T_I,
    \end{dcases}
\eeq
where $Y_{\rm FI}^\text{inst}$ is the dark matter yield produced after reheating, given by Eq.~\eqref{eq:Y0}. The two limits of Eq.~(\ref{eq:YFI}) correspond to the dark matter mass being much larger or smaller than $T_I$. In both cases, dark matter  production  during reheating is exponentially enhanced by a factor $e^{2m/\Trh}$ with respect to the production after reheating. 

The opposite behavior to above occurs for the (atypical) case of $\alpha < 0$,
\beq
    Y_{\rm FI}^\text{rh} \simeq Y_{\rm FI}^\text{inst} \times
    \begin{dcases}
        \frac{1}{|\alpha|} & \text{for } T_I < \Trh \ll m ,\\
        \frac{1}{\alpha   \eta}  \frac{2  m}{\Trh}  e^{\frac{2 m}{\Trh}} & \text{for } T_I < m \ll \Trh ,
    \end{dcases}
\eeq
which corresponds to a subdominant dark matter  production during reheating. Finally, in the (unusual) case that the temperature is constant during reheating, then $\alpha = 0$ \cite{Co:2020xaf, Barman:2022tzk, Chowdhury:2023jft, Cosme:2024ndc} (in which case $T_I = \Tmax = \Trh$) and the yield from reheating is
\beq
    Y_{\rm FI}^\text{rh} &\simeq \frac{135}{128 \pi^7}  \frac{\mathcal K^{q\bar q'}_{QQ'}}{\gss  (3+\omega)}  \sqrt{\frac{10}{\gs}}  \frac{\Mpl}{\Trh}  e^{-\frac{2  m}{\Trh}}\\
    & \qquad \times \left[1 - \left(\frac{a_I}{\arh}\right)^\frac{3(3+\omega)}{2}\right].
\eeq
Thus, parametrically 
\beq
    Y_{\rm FI}^\text{rh}  &\simeq \frac{4}{3(3+\omega)}  \frac{m}{\Trh}  Y_{\rm FI}^\text{inst}.
\eeq
This corresponds to a modest amplification with respect to the production after reheating. We emphasize that the {\it total} dark matter yield corresponds to the production during and after reheating, and therefore the two contributions have to be added
\begin{equation}
    Y_{\rm FI} = Y_{\rm FI}^\text{rh} + Y_{\rm FI}^\text{inst}.
\end{equation}

\begin{figure}[t!]
    \def\sepf{0.99}
    \centering
\includegraphics[width=\sepf\columnwidth]{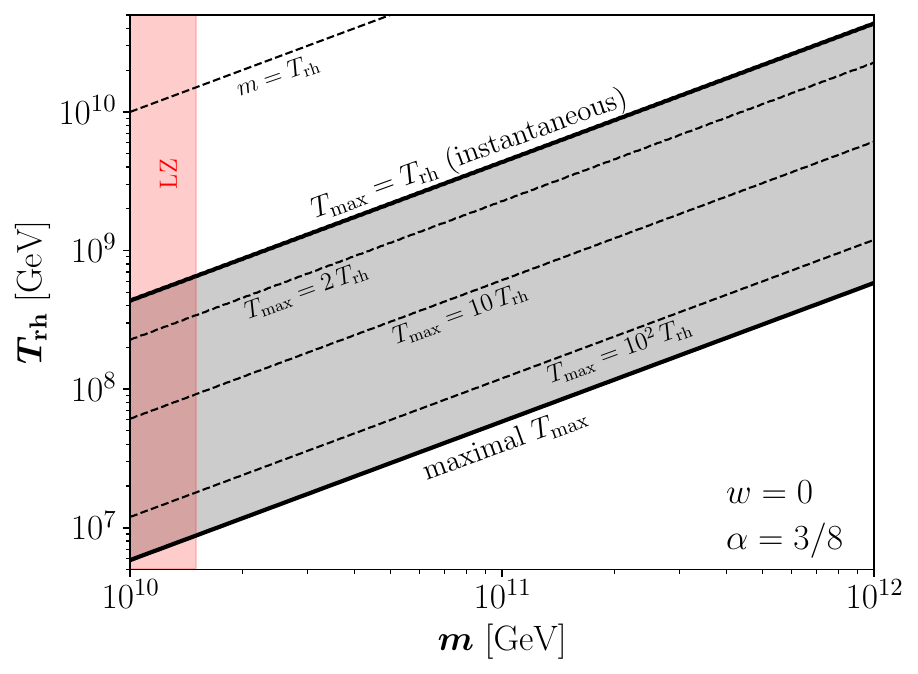}
    \caption{Zoomed-in view of the top left panel of Figure~\ref{fig:non-instantaneous}, overlaying contours corresponding to different values of $\Tmax$.}
    \label{fig:zoom}
\end{figure} 
\begin{figure*}[t!]
    \def\sepf{0.94}
    \centering
    \includegraphics[width=\sepf\columnwidth]{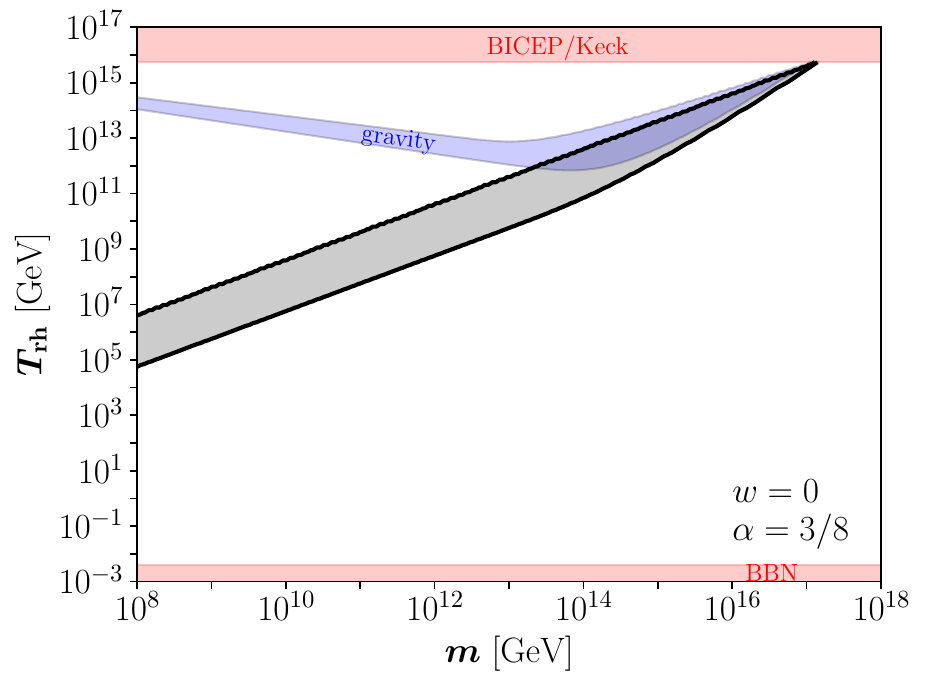}
    \includegraphics[width=\sepf\columnwidth]{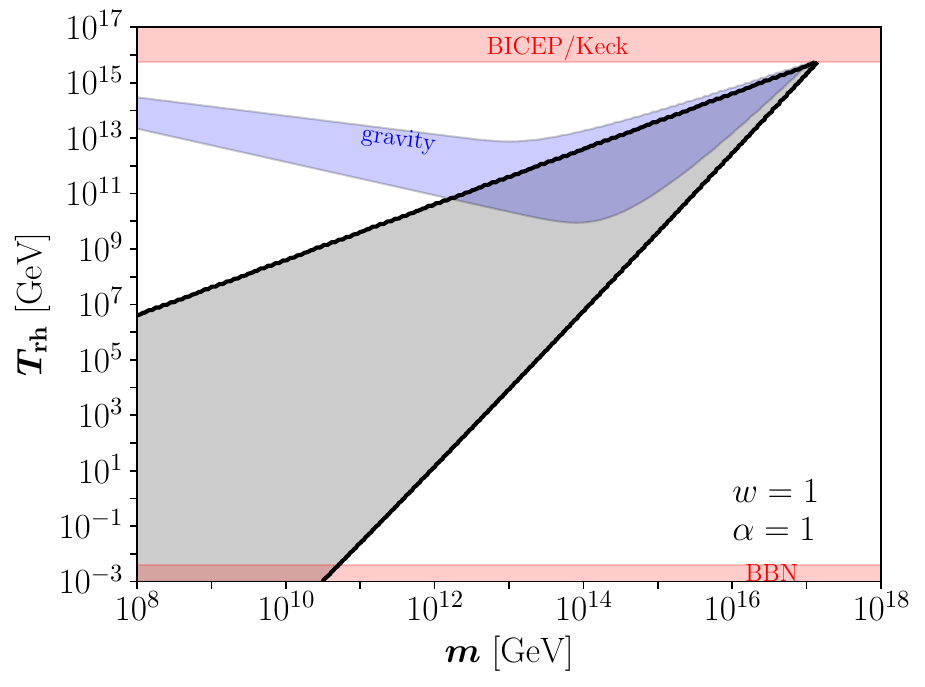}
    \includegraphics[width=\sepf\columnwidth]{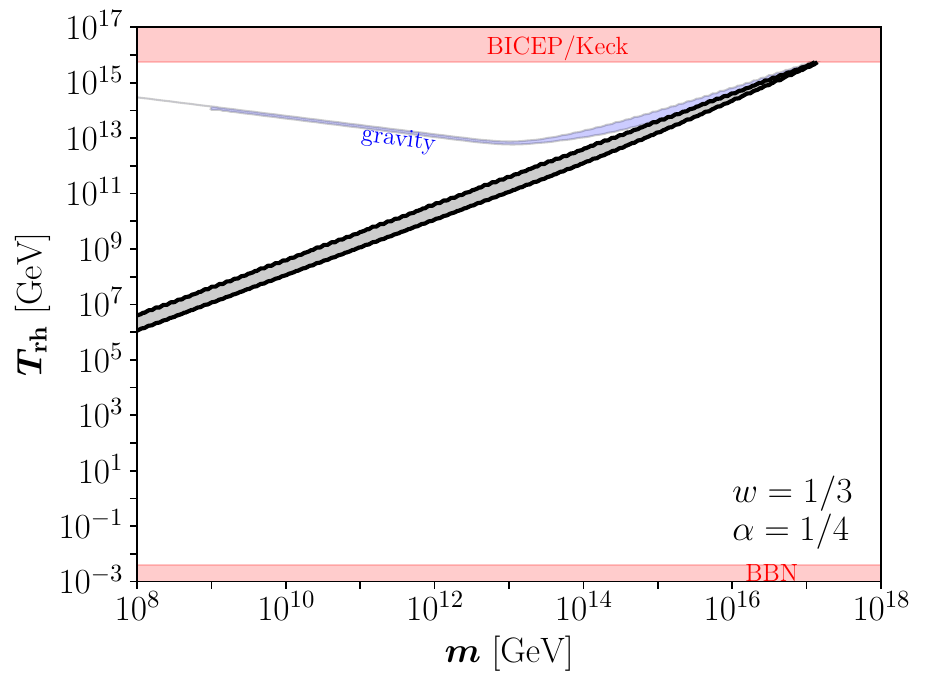}
    \includegraphics[width=\sepf\columnwidth]{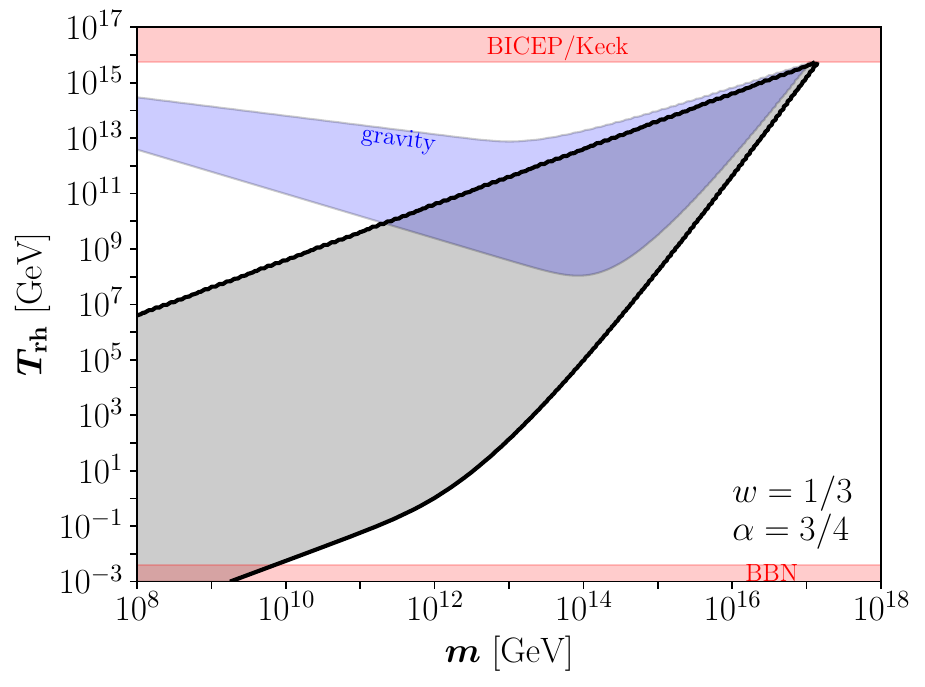}
    \vspace{-2mm}
    \caption{Same as Figure~\ref{fig:non-instantaneous}, overlaying in blue the gravitational production.}
\vspace{-3mm}    \label{fig:gravity}
\end{figure*} 

In Figure~\ref{fig:non-instantaneous} we indicate how non-standard cosmology impacts the expectations for freeze-in of the doublet model outlined in Section \ref{doublet}. We consider different choices of $\omega$ and $\alpha$: the top left panel corresponds to an early matter dominated era ($\omega= 0$, $\alpha = 3/8$), the top right to kination ($\omega= \alpha = 1$), and the lower panels to two versions of a radiation-dominated era ($\omega=1/3$ and $\alpha = 1/4$ or $\alpha = 3/4$). The gray regions correspond to the parameter space where the whole dark matter abundance can be fitted, by choosing $\Tmax$ appropriately, as illustrated in Figure \ref{fig:zoom} (for the case $\omega=0$ with $\alpha=3/8$). Thus this gray area brackets the uncertainty on the duration of the reheating era.  

The upper boundary of the gray areas correspond to instantaneous reheating (that is, $T_I = \Trh$). The lower bounds correspond to the maximum obtainable value of $\Tmax$. In particular, the bound on the inflationary scale $H_I$ in Eq.~\eqref{eq:HI} implies that the duration of reheating has to decrease if $\Trh$ grows, and therefore the bands shrink and collapse to a point when $\Trh \simeq 6 \times 10^{15}$~GeV. The two slopes of the lower bounds reflect the cases in which $m$ is higher and lower than the maximal temperature $T_I$ reached by the thermal bath. The thickness of the gray regions is related to the exponent $\eta - 1$ in Eq.~\eqref{eq:YFI}, with wider bands corresponding to smaller values of $\eta - 1$.
 
Additionally, Figure~\ref{fig:non-instantaneous} is also overlaid with the region in which dark matter thermalizes with the Standard Model, as well as the bounds from direct detection for the doublet representation.

\section{Gravitational Production}
\label{sec6}
Beyond the electroweak interactions discussed previously, dark matter and other heavier states are also inevitably produced via the annihilation of Standard Model particle pairs through $s$-channel graviton exchange. The squared amplitudes for fermionic dark matter are presented in Appendix~\ref{ApB} (cf.~Refs.~\cite{Bernal:2018qlk, Dutra:2019gqz, Clery:2021bwz}).

Under the assumption of Maxwell–Boltzmann statistics, the corresponding reaction density reads
\beq \label{eq:gammagrav}
    \gamma^h_{Q}(T) &= \frac{T^4}{3840 \pi^5  \Mpl^4} \Big(525  m^4 + 1536  m^3  T + 2440  m^2  T^2 \\
    &\quad\quad + 2508  m  T^3 + 1254  T^4\Big)  e^{-\frac{2m}{T}}.
\eeq
In the ultra-relativistic and non-relativistic limits, $\gamma_{Q}$ is
\beq
    \gamma^h_{Q}(T) \simeq
    \begin{dcases}
        \frac{209  T^8}{640 \pi^5  \Mpl^4} &\text{ for } m \ll T ,\\
        \frac{35  m^4  T^4}{256 \pi^5  \Mpl^4}  e^{-\frac{2m}{T}} &\text{ for } T \ll m .
    \end{dcases}
\eeq
The ultra-relativistic limit was reported \cite{Bernal:2018qlk};
the results are compatible, with a difference of approximately 0.5\%.

In the instantaneous reheating approximation, the solution of Eq.~\eqref{eq:yield} yields
\beq \label{eq:yieldgrav}
    Y_\text{FI} = \frac{9}{1024 \pi^8  \gss}  \sqrt{\frac{10}{\gs}}  \frac{f(m,\Trh)}{\Mpl^3}  e^{-\frac{2  m}{\Trh}},
\eeq
where
\beq\notag
    f(m,\Trh) \equiv 525  m^3 + 1536  m^2  \Trh + 1672  m  \Trh^2 + 836  \Trh^3.
\eeq
In the ultra-relativistic and non-relativistic limits, Eq.~\eqref{eq:yieldgrav} reduces to
\beq\notag
    Y_\text{FI}&\simeq
    \begin{dcases}
        \frac{1881}{256 \pi^8  \gss} \sqrt{\frac{10}{\gs}} \left(\frac{\Trh}{\Mpl}\right)^3 &\text{for } m \ll \Trh,\\
        \frac{4725}{1024 \pi^8  \gss}  \sqrt{\frac{10}{\gs}} \left(\frac{m}{\Mpl}\right)^3  e^{-\frac{2 m}{\Trh}} &\text{for } \Trh \ll m.
    \end{dcases}
\eeq

The general case beyond the instantaneous reheating approximation can be computed straightforwardly; however, the resulting expressions are rather lengthy and provide limited additional physical insight. We therefore report them only in Appendix~\ref{ApB}, see Eq.~\eqref{eq:YFIrh}.

Figure~\ref{fig:gravity} is analogous to Figure~\ref{fig:non-instantaneous}, with the gravitational production channel overlaid in blue. The blue regions indicate the parameter space where the observed dark matter abundance can be generated solely through gravitational production. Again, the thicknesses of the regions bracket the uncertainty on the duration of the reheating era. In addition, owing to the strong Planck-mass suppression of this mechanism, sizable dark matter masses and large reheating temperatures are required to achieve the correct relic density. Since this necessitates temperatures higher than those relevant for the electroweak channel, gravitational production never dominates electroweak production in the present scenario.

The dark matter yield from gravitational production is insensitive to the representation. A mild dependence arises, from heavier states that subsequently decay into dark matter. This additional contribution modifies the yield only by an $\mathcal{O}(1)$ factor, which is not discernible over the wide parameter range shown in Figure~\ref{fig:gravity}.

Before closing this section, we note as an aside, that gravitational production is particularly relevant for the singlet fermion representation, as it constitutes the only available production channel, since singlets do not have tree-level couplings to the electroweak gauge bosons.

\vspace{-3mm}\section{Summary and Conclusions} 
\label{sec7}
\vspace{-3mm}
Freeze-in in the regime $m> \Tmax$ offers the prospect of interesting models of dark matter which are highly distinct from {\em both} orthodox freeze-in, and the classic freeze-out WIMP picture, and with the benefit of being potentially discoverable at near-future direct and indirect experiments. Such Boltzmann suppressed freeze-in models,\footnote{Also  called `freeze-in at stronger coupling' \cite{Cosme:2023xpa}, we prefer Boltzmann suppressed freeze-in, as it better includes UV freeze-in \cite{Bernal:2025osg}.}  first outlined in Ref.~\cite{Giudice:2000ex}, have recently been attracting renewed attention \cite{Cosme:2023xpa,  Cosme:2024ndc, Okada:2021uqk, Koivunen:2024vhr, Arcadi:2024wwg, Boddy:2024vgt, Arcadi:2024obp, Bernal:2024ndy, Lee:2024wes, Belanger:2024yoj, Khan:2025keb, Bernal:2025szh, Bernal:2025qkj, Bernal:2025osg, Bernal:2026clv, Feiteira:2026qme}. 

\vspace{-2mm}
\subsection{Summary}
\vspace{-2mm}

In a recent letter \cite{Bernal:2026clv}, we outlined the prospect of the neutral component of a pair of electroweak fermion doublets being dark matter with their relic abundance set via Boltzmann suppressed freeze-in. Due to its extreme minimality we argued that this scenario was worthy of the title: {\em Minimal Freeze-in Dark Matter}. In this companion paper, we have presented the Next-to-Minimal variants, both in terms of field content and cosmology. 

Specifically, we have studied dark matter that arises from different representations of SU(2)${}_L$ focusing on the triplet, quintuplet, and septuplet. While the relic density curves are largely unchanged between the small $n$ representations (see~Figure \ref{fig:instantaneous}),  large differences can arise in expectations for direct and indirect detection (see~Figure~\ref{fig:DD}).
We note, from a field content perspective, the $\mathbf{3}$ with hypercharge $Y=0$ (or any odd $n$ with $Y=0$) could arguably be considered more minimal than the doublet model of  \cite{Bernal:2026clv}, as this provides an anomaly-free extension with just a single Weyl fermion (cf.~Footnote \ref{fn3}). As a counterpoint, there is a compelling simplicity in models with matter only in the fundamental representation.

On the cosmology side, we have explored the implications of deviating from the instantaneous reheating approximation. Moreover, we have highlighted that for non-instantaneous reheating the details of the equation of state of the Universe prior to reheating can have a significant impact on the freeze-in dynamics. Inspection of Figure \ref{fig:non-instantaneous} indicates the wider parameter freedom that non-minimal cosmological assumptions permit.

\vspace{-2mm}
\subsection{Scalar Variants}
\vspace{-2mm}

We have endeavored to explore the cleanest and best motivated examples of Boltzmann suppressed freeze-in electroweak dark matter. In particular, we have restricted our considerations to fermion dark matter. The reasons  we favor fermions over scalars are two-fold: $(i).$~A second scalar implies a second hierarchy problem, $(ii).$~Naturalness suggests that generically the operator $\kappa|\phi|^2|H|^2$ will appear with $\kappa\sim\mathcal{O}(1)$. The latter point is not so much a problem, rather it makes the analysis less clean as it introduces another free parameter.

Putting aside naturalness considerations, one might consider dark matter, which is a complex scalar $\phi$ transforming as a non-trivial representation  under SU(2)${}_L$. Let us briefly discuss the case in which $\phi$ is a scalar doublet. This is reminiscent of a second Higgs without Yukawa couplings. Similarly to the fermion case, for an SU(2)${}_L$ doublet complex scalar $\phi$ with hypercharge $Y$, for one of the components to be neutral, the hypercharge must be $Y=\pm1/2$. The Lagrangian contribution corresponding to the scalar electroweak doublet is given by
\beq
    \Delta\mathcal L = (D_\mu \phi)^\dagger (D^\mu \phi) - m_\phi^2 |\phi|^2 +\kappa|\phi|^2|H|^2
\eeq
where the above assumes a $\mathbb{Z}_2$ symmetry $\phi\rightarrow-\phi$. To make the model predictive, one must assume $\kappa\ll 1$ (although see Refs.~\cite{Cosme:2023xpa, Bernal:2025fcl} for studies of the Higgs portal). If the Higgs portal is negligible (which is an assumption which goes against technical naturalness), the leading interactions are via the electroweak gauge bosons, which interestingly leads to derivative couplings. Thus, the scalar dark matter variant is a non-trivial extension of the models considered here, as such we leave it for future studies.

\vspace{-2mm}
\subsection{Motivations for Higher Representations}
\vspace{-2mm}

Electroweak dark matter is independently motivated from a UV perspective. Notably, in the context of supersymmetry \cite{Martin:1997ns}, Higgsinos and Winos provide quintessential examples of doublet and triplet dark matter (stabilized by R-parity). It is natural for the Higgsino to be the lightest supersymmetric particle (LSP) in scenarios involving the Giudice-Masiero mechanism \cite{Giudice:1988yz}, whereas gauginos such as the Wino are naturally the LSP in R-symmetric models, e.g.~\cite{Fayet:1977yc, Farrar:1982te, Hall:1990hq, Fox:2002bu, Kribs:2007ac, Unwin:2024yqq, Unwin:2012fj}.
Moreover, it has been suggested that in supersymmetric models the inclusion of  $\mathbf{5}$ can resolve the `little hierarchy problem' if the Standard Model superpartners are pushed to $\sim$10 TeV \cite{Fabbrichesi:2015bta}. The LSP dark matter in this model is the neutral fermion component of the $\mathbf{5}$, and thus is similar to the scenario considered here.

Aside from supersymmetry, alternative motivations have been found, especially for the $\mathbf{5}$ representation. For instance, the quintuplet has been proposed to  have a possible origin in SU(5) GUTs \cite{Toma:2024lmt}, as well as potentially arising in extra dimensional settings, in the context of 5D gauge-Higgs unification \cite{Maru:2017pwl}.

It should also be emphasised that the automatic metastability of dark matter arising from the $\mathbf{5}$  and $\mathbf{7}$ representations is a highly desirable feature. Not only does metastability remove the need for an {\em ad hoc} stabilizing symmetry,  it also avoids the question of UV completion. We note that if a global symmetry (e.g.~$\mathbb{Z}_2$ or U(1)) stabilizes the dark matter, then Planck induced operators are generically expected to violate this symmetry \cite{Banks:2010zn}. Accordingly, dimension five Planck-suppressed operators will commonly introduce dark matter decays unless these operators are forbidden by UV completing the global symmetry into a local symmetry or if the dark matter is accidentally metastable. This issue can be evaded if dark matter arises from a $\mathbf{5}$ or $\mathbf{7}$ of SU(2)${}_L$, since dark matter can be accidentally metastable by construction.

\vspace{-2mm}
\subsection{Closing Remarks}
\vspace{-2mm}

Forthcoming and proposed experiments offer the potential to discover this class of dark matter particles. Notably, direct and indirect detection already constrain the parameter space of these models, and near-future observations will continue to advance the search for these states. In particular, DARWIN \cite{DARWIN:2016hyl} and CTAO \cite{Garcia-Cely:2015dda, Abe:2025lci} (for direct and indirect detection, respectively) improve on the reach of current experiments by orders-of-magnitude \cite{Das:2024bed}.

Arguably, the most attractive variants are those that are automatically metastable.
It is notable that in contrast to the regular TeV scale freeze-out Minimal Dark Matter, heavy dark matter coming from the quintuplet representation can lead to decay signals that may be detectable over the lifetime of KM3NeT \cite{Ng:2020ghe, Kohri:2025bsn}. We also highlight the tentative claim that a PeV neutrino event may be  indicative of decaying dark matter \cite{Kohri:2025bsn}. As seen in Section \ref{sec4}, decaying dark matter is consistent with $\Psi^0$ arising from the (viable) $\mathbf{5}$ model with $Y=0$.

Although the collider bounds do not currently provide competitive constraints \cite{Panci:2024oqc, Ostdiek:2015aga}, future colliders could constrain (or discover) $\sim$ TeV SU(2)${}_L$ multiplets. Studies have been made of minimal dark matter searches using proposed colliders based on linear $e^+e^-$ \cite{Kumar:2021umc}, 100~TeV $pp$ \cite{Zeng:2019tlw}, and muon colliders \cite{Bottaro:2021srh}. In particular, this Next-to-Minimal freeze-in scenario appears to be an ideal benchmark for any future wakefield collider; see, e.g.~\cite{Gessner:2025acq, Chigusa:2025azs}.

The models outlined above are extremely predictive since the dark matter phenomenology is entirely determined by the dark matter mass for a given representation (since the couplings are fixed to be electroweak). 
Unlike conventional freeze-in which is coupled extremely weakly to the Standard Model, the models presented here offer the potential for discovery in forthcoming experiments and provide excellent benchmarks for future searches. 

\vspace{2mm}
{\bf Acknowledgments.}~We thank Qing Chen, Camilo García-Cely, Richard~Hill, and Marta~Losada for helpful interactions. NB received grants PID2023-151418NB-I00 funded by MCIU/AEI/10.13039/501100011033/FEDER and PID2022-139841NB-I00  of MICIU/AEI/10.13039/ 501100011033 and FEDER, UE. JU is supported by NSF grant PHY-2209998. 

\appendix
\vspace{-2mm}
\section{Production cross-section}
\label{ApA}
\vspace{-2mm}
Below we derive the production cross-sections: 
\beq\notag
q\bar q' \to \Psi^+ \Psi^-,~ \Psi^0 \bar\Psi^0,~  \Psi^\pm\bar\Psi^0,~  \Psi^- \Psi^{++},~ \Psi^{--} \Psi^{+++},\cdots
\eeq
This section largely follows the appendix of \cite{Bernal:2026clv} with a few generalizations.
The process $  q(p_1)  \bar q'(p_2) \to \Psi(k_1)  \Psi'(k_2)$ leading to the production of a pair of fermions with mass $m$ and momenta $k_1,k_2$ from two quarks with momenta $p_1,p_2$ via a vector boson arises from the Lagrangian
\beq\label{L}
\notag
  \mathcal{L}_{\mathrm{int}} = V_\mu\left[\bar q \gamma^\mu\left(g_V^{q\bar q'} - g_A^{q\bar q'} \gamma^5\right) q' + \bar \Psi \gamma^\mu\left(g_V^\Psi - g_A^\Psi \gamma^5\right)\Psi'\right],
\eeq
with generic vector couplings  $g_V^{q\bar q'},~g_V^{\Psi}$ and axial-vector couplings $g_A^{q\bar q'},~g_A^{\Psi}$, as given in Eq.~\eqref{qqq}. 

Neglecting the mediator mass, the propagator is $- i/s$  the tree-level amplitude is
\beq
  \mathcal{M} = \frac{-i}{s}& \big[\bar v(p_2) \gamma^\mu \left(g_V^{q\bar q'} - g_A^{q\bar q'} \gamma^5\right) u(p_1)\big]\\
  &\times\big[\bar u(k_1) \gamma_\mu \left(g_V^\Psi - g_A^\Psi \gamma^5\right) v(k_2)\big].
\eeq
In the following, we neglect quark masses and take the limit $m_Z^2\ll m^2$, $s$.
We square the matrix element and sum over the final-state colors and spins, averaging over initial-state colors and spins, it follows 
\beq
    \overline{|\mathcal{M}|^2} = \frac{1}{12 s^2} & {\rm Tr} \Big[ \slashed{p}_2 \gamma^\mu \left(g_V^{q\bar q'} - g_A^{q\bar q'} \gamma^5\right) \slashed{p}_1 \gamma^\nu \left(g_V^{q\bar q'} - g_A^{q\bar q'} \gamma^5\right) \Big] \\
    &\times{\rm Tr} \Big[ (\slashed{k}_1 + m)\gamma_\mu \left(g_V^\Psi - g_A^\Psi \gamma^5\right)\\ &~~\qquad\cdot(\slashed{k}_2 - m)\gamma_\nu \left(g_V^\Psi - g_A^\Psi \gamma^5\right) \Big].
    \eeq
 We can re-express this in terms of the Mandelstam invariants,
\beq
    s=&(p_1+p_2)^2,\\
    t=&(p_1-k_1)^2,\\
    u=&(p_1-k_2)^2,
\eeq
with $s+t+u=2m^2$ for massless initial-state quarks. Accordingly, one has   
\beq
 \overline{|\mathcal{M}|^2}= \frac{2}{3 s^2}\Bigg[ &\Big((g_V^{q})^2+(g_A^{q})^2\Big) \Big( (g_V^{\Psi})^2\big(t^2+u^2+2m^2 s\big) \\
 &\qquad + (g_A^{\Psi})^2\big(t^2+u^2-2m^2 s\big) \Big)\\
 &\qquad +4  g_V^{q\bar q'} g_A^{q\bar q'} g_V^\Psi g_A^\Psi  s (t-u)\Bigg].
    \label{eq:2}
\eeq
Defining
$    \beta \equiv \sqrt{1-\frac{4m^2}{s}}$,
and $\vartheta$ as the angle between $\vec p_1$ and $\vec k_1$, one has
\beq
    t &= m^2-\frac{s}{2}\left(1-\beta\cos\vartheta\right),\\
    u &= m^2-\frac{s}{2}\left(1+\beta\cos\vartheta\right).
\eeq

The differential cross-section is
\beq
    \frac{d\sigma}{d\Omega} = \frac{1}{64 \pi^2  s}   \frac{|\vec k|}{|\vec p|}  \overline{|\mathcal{M}|^2} = \frac{\beta}{64 \pi^2  s}  \overline{|\mathcal{M}|^2} .
\eeq
Substituting Eq.~\eqref{eq:2} and integrating the azimuthal angle to get a factor $2\pi$ gives
\vspace*{3mm}
\begin{widetext}
\beq\notag
    \frac{d\sigma}{d\cos\vartheta} &= \frac{\beta}{32\pi  s}  \frac{1}{3} \Bigg\{ \Big((g_V^{q\bar q'})^2 + (g_A^{q\bar q'})^2\Big)\Big[(g_V^{\Psi})^2 \left(1+\cos^2\vartheta + (1-\beta^2)\sin^2\vartheta\right) + (g_A^\Psi)^2 \beta^2(1+\cos^2\vartheta)\Big]
   + 4  g_V^{q\bar q'}   g_A^{q\bar q'}  g_V^\Psi  g_A^\Psi \beta\cos\vartheta \Bigg\},
\eeq
\end{widetext}
where we used $m^2 = \frac{s}{4}(1-\beta^2)$ and have integrated the azimuthal angle over $2\pi$. Integrating the differential cross-section over $\cos\vartheta \in [-1,1]$ leads to the cross-section for $q\bar q\to \Psi\Psi$ given by
\beq\notag\small
    \sigma
&    = \frac{\beta}{24\pi s} \Big[(g_V^{q\bar q'})^2 + (g_A^{q\bar q'})^2\Big] \left( (g_V^{\Psi})^2- \frac{\beta^2}{3}\left((g_V^{\Psi})^2 -  2(g_A^{\Psi})^2  \right)\right)
    \label{yu}
\eeq
Note that when integrating over $\cos\vartheta$ the term proportional to $g_V^{q\bar q'}  g_A^{q\bar q'}  g_V^\Psi  g_A^\Psi$ vanishes.
We re-express this in the following manner
\beq
    \sigma_{q\bar q'\to \Psi_Q\bar\Psi_{Q'}}(\beta) \simeq \frac{\beta}{24\pi s}  \mathcal{K}^{q\bar q'}_{Q,Q'} ,
    \label{eq:xsec2}
\eeq
with $\mathcal{K}^{q\bar q'}_{Q,Q'}$ defined (as in eq.~(\ref{9})) to be
\beq 
    \mathcal{K}^{q\bar q'}_{QQ'} \equiv \delta_{qq'} \left(\mathcal{C}^{q\bar q}_{Q,\gamma} + \mathcal{C}^{q\bar q}_{Q,Z}\right) + (1 - \delta_{qq'})\mathcal{C}^{q\bar q'}_{QQ',W}
\eeq
Recall, at leading order (cf.~Eq.~(\ref{11})) for  $i=\gamma,Z$
\beq
   \mathcal C^{q\bar q'}_{Q,i} \approx \Big( (g_{V}^{q\bar q',i})^2 +  (g_{A}^{q\bar q',i})^2\Big)  (g^{\Psi,i}_{V,Q})^2.
\eeq
For the photon channel ($q=q'$) the quark axial coupling vanishes $g_A^{q,\gamma}=0$, the vector coupling is $g_V^{q,\gamma}=eQ_q$, and the $\mathcal{T}^{\rm th}$ component coupling is $g^{\Psi,\gamma}_{V,Q}=eQ$, hence
\beq
    \mathcal C^{q}_{Q,\gamma}=e^4 Q_q^2 Q^2.
\eeq

For the $Z$ channel ($q=q'$) the quark couplings are the Standard Model ones (cf.~Eq.~\eqref{q}), and recall
\beq
    g^{\Psi,Z}_{V,Q}=\frac{e}{s_W c_W}\Big(\mathcal{T}-Q s_W^2\Big),
\eeq
hence for the isospin component $\mathcal{T}$ with charge $Q$ one has
\beq
    \mathcal C^{q}_{Q,Z}
   = \Big((g^{q,Z}_{V})^2+(g^{q,Z}_{A})^2\Big) 
    \left[\frac{e}{s_W c_W}\Big(\mathcal{T}-Q s_W^2\Big)\right]^2.
\eeq

$W^\pm$ induced `co-production' processes, such as
\beq
    u(p_1) \bar d(p_2)\to W^+ & \to \Psi^+(k_1) \bar\Psi^0(k_2),
    \\[5pt]
    d(p_1) \bar u(p_2)\to W^- & \to \Psi^-(k_1) \Psi^0(k_2),
\eeq
the cross-section in the limit $\beta\ll1$ is of the form
\beq
    \sigma^{(W)}_{q\bar q'\to \Psi_Q\bar\Psi_{Q-1}} \Big|_{\beta\ll1} \simeq \frac{\beta}{24\pi s} \mathcal C_{Q,W}^{q\bar q'}.
\eeq
Equation~\eqref{eq:xsec2} can be specialized to this case by using couplings of the form (for $q\neq q'$)
\beq
g_{V,W}^{q\bar q'} &=g_{A,W}^{q\bar q'}=g_{V,W}^{q'\bar q}=g_{A,W}^{q'\bar q}=\frac{e}{2\sqrt{2}s_W} V_{qq'}, \\[5pt] g^{\Psi,W}_{V,Q} &=g^{\Psi,W}_{A,Q}=\frac{e}{2\sqrt{2} s_W}\sqrt{(T+\mathcal{T})(T-\mathcal{T}+1)},
\eeq
yielding
\beq
    \mathcal C^{u_i d_j}_{Q,W}
    = \frac{1}{32}\frac{e^4}{s_W^4} |V_{ij}|^2 (T+\mathcal{T})(T-\mathcal{T}+1).
\eeq
In the above, we have worked in the $(\gamma,Z)$ basis and the $\Psi_{Q}$ fields, since direct detection limits require freeze-in to occur at temperatures $T\gg m_Z$ it may be more intuitive to work in the unbroken basis (e.g.~in terms of $\chi$ fields and basis $(B,W_3)$). Since at a late-time we wish to identify the relic abundance of the dark matter state $\Psi^0$, it seems simpler to always work in the basis of the broken phase. 

\vspace*{5mm}
\section{Graviton exchange}
\label{ApB}
The total amplitude squared for graviton exchange can be expressed as the sum of three contributions, weighted by the Standard Model content in fields:
\begin{equation}
    |\mathcal{M}|^2 = 4  |\mathcal{M}_0|^2 + 45  |\mathcal{M}_\frac12|^2 + 12  |\mathcal{M}_1|^2.
\end{equation}
Here we give the squared amplitudes for fermionic dark matter (where we indicate the spin of the initial-state particle in the subscript) \cite{Bernal:2018qlk, Dutra:2019gqz, Clery:2021bwz}
\begin{widetext}
\beq
    |\mathcal{M}_0|^2 &= \frac{1}{32  \Mpl^4  s^2} \left(m^2 - t\right) \left(s + t - m^2\right) \left(s + 2t - 2m^2\right)^2 , \\[8pt]
    |\mathcal{M}_\frac12|^2 &= \frac{1}{128  \Mpl^4  s^2} \Big[32  m^8 - 32  m^6 (s + 4 t) + 2  m^4 (5  s^2 + 64  s  t + 96  t^2) \\&\hspace{2.5cm}
    - 4  m^2 (s^3 + 13  s^2  t + 40  s  t^2 + 32  t^3) + s^4 + 10  s^3  t + 42  s^2  t^2 + 64  s  t^3 + 32  t^4\Big], \\[8pt]
    |\mathcal{M}_1|^2 &= \frac{-1}{8  \Mpl^4  s^2} \left[m^4 - 2  m^2  t + t  (s + t)\right] \left[2 \left(m^4 - 2  m^2  t + t  (s + t)\right) + s^2\right].
\eeq
\end{widetext}

Then the dark matter production rate density $\gamma^h_{Q}$ is \cite{Gondolo:1990dk}
\begin{widetext}
    \begin{equation}
        \gamma^h_{Q}(T) = \frac{1}{4 (2\pi)^4} \int_m^\infty dE_1 dE_2  e^{-\frac{E_1}{T}} e^{-\frac{E_2}{T}}\int_{4m^2}^{4E_1E_2} ds \frac{s}{\sqrt{\lambda(s,m^2,m^2)}} \int_{t_-}^{t_+}dt  \frac{|\mathcal{M}|^2}{8\pi  s} ,
    \end{equation}
\end{widetext}
where
\begin{align}
    &\lambda(s,m^2,m^2) = s  (s - 4  m^2) ,\\[5pt]
    &t_- = - \frac{1}{4  s} \left(s + \sqrt{s  (s - 4m^2)}\right)^2,\\[5pt]
    &t_+ = m^2 + \frac12 \left(-s + \sqrt{s  (s - 4m^2)}\right),
\end{align}
one gets
\beq
    \gamma^h_{Q}(T) &= \frac{T^4}{3840 \pi^5  \Mpl^4} \Big(525  m^4 + 1536  m^3  T \\
    &\quad + 2440  m^2  T^2 + 2508  m  T^3 + 1254  T^4\Big)  e^{-\frac{2m}{T}},
\eeq
which corresponds to Eq.~\eqref{eq:gammagrav}.

Before closing, we report the dark matter yield produced during reheating, which corresponds to the result of Eq.~\eqref{eq:dNda} with $\gamma^h_{Q}$. One gets
\begin{align} \label{eq:YFIrh}
    Y_\text{FI}^\text{rh} &= \frac{9}{2^{10} \pi^8  \gss} \sqrt{\frac{10}{\gs}} \left(\frac{m}{\Mpl}\right)^3 \left(\frac{\Trh}{2m}\right)^{q-2}\nonumber\\
    &\quad\times\Big[525  \tilde\Gamma(q-1) + 3072   \tilde\Gamma(q-2) + 9760  \tilde\Gamma(q-3) \nonumber\\
    &\qquad+ 20064 \left(\tilde\Gamma(q-4) + \tilde\Gamma(q-5)\right) \Big],
\end{align}
with
\beq
    \tilde\Gamma(x) &\equiv \Gamma\left(x, \frac{2m}{\Trh}\right) - \Gamma\left(x, \frac{2m}{T_I}\right),\\
    q &\equiv \frac{3  (1+\omega)}{2 \alpha} .
\eeq

\bibliographystyle{JHEP} 
\bibliography{biblio}

@article{Clery:2021bwz,
    author = "Cléry, Simon and Mambrini, Yann and Olive, Keith A. and Verner, Sarunas",
    title = "{Gravitational portals in the early Universe}",
    eprint = "2112.15214",
    archivePrefix = "arXiv",
    primaryClass = "hep-ph",
    reportNumber = "UMN-TH-4110/22, FTPI-MINN-22/02, CERN-TH-2021-222",
    doi = "10.1103/PhysRevD.105.075005",
    journal = "Phys. Rev. D",
    volume = "105",
    number = "7",
    pages = "075005",
    year = "2022"
}

@article{Deligny:2024fyx,
    author = "Deligny, Olivier",
    title = "{Constraints on superheavy dark matter decaying into $h\nu $, $Z\nu $ and $W\ell $}",
    eprint = "2408.17111",
    archivePrefix = "arXiv",
    primaryClass = "hep-ph",
    doi = "10.1140/epjc/s10052-025-14736-3",
    journal = "Eur. Phys. J. C",
    volume = "85",
    number = "9",
    pages = "985",
    year = "2025"
}

@article{Das:2023wtk,
    author = "Das, Saikat and Murase, Kohta and Fujii, Toshihiro",
    title = "{Revisiting ultrahigh-energy constraints on decaying superheavy dark matter}",
    eprint = "2302.02993",
    archivePrefix = "arXiv",
    primaryClass = "astro-ph.HE",
    reportNumber = "YITP-23-08",
    doi = "10.1103/PhysRevD.107.103013",
    journal = "Phys. Rev. D",
    volume = "107",
    number = "10",
    pages = "103013",
    year = "2023"
}

@article{Gondolo:1990dk,
    author = "Gondolo, Paolo and Gelmini, Graciela",
    title = "{Cosmic abundances of stable particles: Improved analysis}",
    reportNumber = "UCLA-90-TEP-68",
    doi = "10.1016/0550-3213(91)90438-4",
    journal = "Nucl. Phys. B",
    volume = "360",
    pages = "145--179",
    year = "1991"
}

@article{Martin:1997ns,
    author = "Martin, Stephen P.",
    editor = "Kane, Gordon L.",
    title = "{A Supersymmetry primer}",
    eprint = "hep-ph/9709356",
    archivePrefix = "arXiv",
    reportNumber = "FERMILAB-PUB-97-425-T",
    doi = "10.1142/9789812839657_0001",
    journal = "Adv. Ser. Direct. High Energy Phys.",
    volume = "18",
    pages = "1--98",
    year = "1998"
}

@article{Kribs:2007ac,
    author = "Kribs, Graham D. and Poppitz, Erich and Weiner, Neal",
    title = "{Flavor in supersymmetry with an extended R-symmetry}",
    eprint = "0712.2039",
    archivePrefix = "arXiv",
    primaryClass = "hep-ph",
    doi = "10.1103/PhysRevD.78.055010",
    journal = "Phys. Rev. D",
    volume = "78",
    pages = "055010",
    year = "2008"
}

@article{Das:2024bed,
    author = "Das, Saikat and Carpio, Jose Alonso and Murase, Kohta",
    title = "{Probing superheavy dark matter through lunar radio observations of ultrahigh-energy neutrinos and the impacts of neutrino cascades}",
    eprint = "2405.06382",
    archivePrefix = "arXiv",
    primaryClass = "hep-ph",
    doi = "10.1103/PhysRevD.111.083007",
    journal = "Phys. Rev. D",
    volume = "111",
    number = "8",
    pages = "083007",
    year = "2025"
}

@article{Farrar:1982te,
    author = "Farrar, Glennys R. and Weinberg, Steven",
    title = "{Supersymmetry at Ordinary Energies. 2. R Invariance, Goldstone Bosons, and Gauge Fermion Masses}",
    reportNumber = "RU-82-38",
    doi = "10.1103/PhysRevD.27.2732",
    journal = "Phys. Rev. D",
    volume = "27",
    pages = "2732",
    year = "1983"
}

@article{Fayet:1977yc,
    author = "Fayet, Pierre",
    title = "{Spontaneously Broken Supersymmetric Theories of Weak, Electromagnetic and Strong Interactions}",
    reportNumber = "LPTENS 77/11",
    doi = "10.1016/0370-2693(77)90852-8",
    journal = "Phys. Lett. B",
    volume = "69",
    pages = "489",
    year = "1977"
}

@article{Hall:1990hq,
    author = "Hall, L. J. and Randall, Lisa",
    title = "{U(1)-R symmetric supersymmetry}",
    reportNumber = "UCB-PTH-90-24, LBL-29561",
    doi = "10.1016/0550-3213(91)90444-3",
    journal = "Nucl. Phys. B",
    volume = "352",
    pages = "289--308",
    year = "1991"
}

@article{Fox:2002bu,
    author = "Fox, Patrick J. and Nelson, Ann E. and Weiner, Neal",
    title = "{Dirac gaugino masses and supersoft supersymmetry breaking}",
    eprint = "hep-ph/0206096",
    archivePrefix = "arXiv",
    reportNumber = "UW-PT-02-12, UW-02-12",
    doi = "10.1088/1126-6708/2002/08/035",
    journal = "JHEP",
    volume = "08",
    pages = "035",
    year = "2002"
}

@article{Unwin:2012fj,
    author = "Unwin, James",
    title = "{R-symmetric High Scale Supersymmetry}",
    eprint = "1210.4936",
    archivePrefix = "arXiv",
    primaryClass = "hep-ph",
    doi = "10.1103/PhysRevD.86.095002",
    journal = "Phys. Rev. D",
    volume = "86",
    pages = "095002",
    year = "2012"
}

@article{Unwin:2024yqq,
    author = "Unwin, James and Yildirim, Tom",
    title = "{A QCD R-axion}",
    eprint = "2407.17557",
    archivePrefix = "arXiv",
    primaryClass = "hep-ph",
    doi = "10.1103/PhysRevD.111.015048",
    journal = "Phys. Rev. D",
    volume = "111",
    number = "1",
    pages = "015048",
    year = "2025"
}

@article{Giudice:1988yz,
    author = "Giudice, G. F. and Masiero, A.",
    title = "{A Natural Solution to the $\mu$ Problem in Supergravity Theories}",
    reportNumber = "SISSA-17/88/EP",
    doi = "10.1016/0370-2693(88)91613-9",
    journal = "Phys. Lett. B",
    volume = "206",
    pages = "480--484",
    year = "1988"
}

@article{Einasto:1965czb,
    author = "Einasto, J.",
    title = "{On the Construction of a Composite Model for the Galaxy and on the Determination of the System of Galactic Parameters}",
    journal = "Trudy Astrofizicheskogo Instituta Alma-Ata",
    volume = "5",
    pages = "87--100",
    year = "1965"
}

@article{Ng:2020ghe,
    author = "Ng, Kenny C. Y. and others",
    title = "{Sensitivities of KM3NeT on decaying dark matter}",
    eprint = "2007.03692",
    archivePrefix = "arXiv",
    primaryClass = "astro-ph.HE",
    month = "7",
    year = "2020"
}

@article{Gessner:2025acq,
    author = "Gessner, Spencer and others",
    title = "{Design Initiative for a 10 TeV pCM Wakefield Collider}",
    eprint = "2503.20214",
    archivePrefix = "arXiv",
    primaryClass = "physics.acc-ph",
    reportNumber = "FERMILAB-CONF-25-0214-CSAID",
    month = "3",
    year = "2025"
}

@article{Abe:2025lci,
    author = "Abe, Shotaro and others",
    title = "{Discovering the Higgsino at CTAO-North within the Decade}",
    eprint = "2506.08084",
    archivePrefix = "arXiv",
    primaryClass = "hep-ph",
    month = "6",
    year = "2025"
}

@article{Panci:2024oqc,
    author = "Panci, Paolo",
    title = "{Electroweak Multiplets as Dark Matter candidates: A brief review}",
    eprint = "2405.05087",
    archivePrefix = "arXiv",
    primaryClass = "hep-ph",
    reportNumber = "C23-04-23",
    doi = "10.22323/1.463.0033",
    journal = "PoS",
    volume = "CORFU2023",
    pages = "033",
    year = "2024"
}

@article{Chigusa:2025azs,
    author = "Chigusa, So and others",
    title = "{Searches for electroweak states at future plasma wakefield colliders}",
    eprint = "2512.09995",
    archivePrefix = "arXiv",
    primaryClass = "hep-ph",
    month = "12",
    year = "2025"
}

@article{HESS:2013rld,
    author = "Abramowski, A. and others",
    collaboration = "H.E.S.S.",
    title = "{Search for Photon-Linelike Signatures from Dark Matter Annihilations with H.E.S.S.}",
    eprint = "1301.1173",
    archivePrefix = "arXiv",
    primaryClass = "astro-ph.HE",
    doi = "10.1103/PhysRevLett.110.041301",
    journal = "Phys. Rev. Lett.",
    volume = "110",
    pages = "041301",
    year = "2013"
}

@article{Feng:2010zp,
    author = "Feng, Jonathan L. and Kaplinghat, Manoj and Yu, Hai-Bo",
    title = "{Sommerfeld Enhancements for Thermal Relic Dark Matter}",
    eprint = "1005.4678",
    archivePrefix = "arXiv",
    primaryClass = "hep-ph",
    reportNumber = "UCI-TR-2010-06",
    doi = "10.1103/PhysRevD.82.083525",
    journal = "Phys. Rev. D",
    volume = "82",
    pages = "083525",
    year = "2010"
}

@article{Banks:2010zn,
    author = "Banks, Tom and Seiberg, Nathan",
    title = "{Symmetries and Strings in Field Theory and Gravity}",
    eprint = "1011.5120",
    archivePrefix = "arXiv",
    primaryClass = "hep-th",
    doi = "10.1103/PhysRevD.83.084019",
    journal = "Phys. Rev. D",
    volume = "83",
    pages = "084019",
    year = "2011"
}

@article{Garcia-Cely:2015dda,
    author = "García-Cely, Camilo and Ibarra, Alejandro and Lamperstorfer, Anna S. and Tytgat, Michel H. G.",
    title = "{Gamma-rays from Heavy Minimal Dark Matter}",
    eprint = "1507.05536",
    archivePrefix = "arXiv",
    primaryClass = "hep-ph",
    doi = "10.1088/1475-7516/2015/10/058",
    journal = "JCAP",
    volume = "10",
    pages = "058",
    year = "2015"
}

@article{Fabbrichesi:2015bta,
    author = "Fabbrichesi, Marco and Urbano, Alfredo",
    title = "{Natural minimal dark matter}",
    eprint = "1510.03861",
    archivePrefix = "arXiv",
    primaryClass = "hep-ph",
    doi = "10.1103/PhysRevD.93.055017",
    journal = "Phys. Rev. D",
    volume = "93",
    number = "5",
    pages = "055017",
    year = "2016"
}

@article{Maru:2017pwl,
    author = "Maru, Nobuhito and Okada, Nobuchika and Okada, Satomi",
    title = "{Fermionic Minimal Dark Matter in 5D Gauge-Higgs Unification}",
    eprint = "1801.00686",
    archivePrefix = "arXiv",
    primaryClass = "hep-ph",
    reportNumber = "OCU-PHYS-470, YGHP17-08",
    doi = "10.1103/PhysRevD.96.115023",
    journal = "Phys. Rev. D",
    volume = "96",
    number = "11",
    pages = "115023",
    year = "2017"
}

@article{Toma:2024lmt,
    author = "Toma, Takashi",
    title = "{Minimal dark matter in SU(5) grand unification}",
    eprint = "2412.19660",
    archivePrefix = "arXiv",
    primaryClass = "hep-ph",
    reportNumber = "KANAZAWA-24-09",
    doi = "10.1103/PhysRevD.111.L051701",
    journal = "Phys. Rev. D",
    volume = "111",
    number = "5",
    pages = "L051701",
    year = "2025"
}

@article{Kohri:2025bsn,
    author = "Kohri, Kazunori and Paul, Partha Kumar and Sahu, Narendra",
    title = "{Super heavy dark matter origin of the PeV neutrino event: KM3-230213A}",
    eprint = "2503.04464",
    archivePrefix = "arXiv",
    primaryClass = "hep-ph",
    reportNumber = "KEK-TH-2708, KEK-Cosmo-0375",
    doi = "10.1103/vvqq-1z2t",
    journal = "Phys. Rev. D",
    volume = "112",
    number = "3",
    pages = "L031703",
    year = "2025"
}

@article{Kumar:2021umc,
    author = "Kumar, Nilanjana and Sahdev, Vandana",
    title = "{Alternative signatures of the quintuplet fermions at the LHC and future linear colliders}",
    eprint = "2112.09451",
    archivePrefix = "arXiv",
    primaryClass = "hep-ph",
    doi = "10.1103/PhysRevD.105.115016",
    journal = "Phys. Rev. D",
    volume = "105",
    number = "11",
    pages = "115016",
    year = "2022"
}

@article{Bottaro:2021srh,
    author = "Bottaro, Salvatore and Strumia, Alessandro and Vignaroli, Natascia",
    title = "{Minimal Dark Matter bound states at future colliders}",
    eprint = "2103.12766",
    archivePrefix = "arXiv",
    primaryClass = "hep-ph",
    doi = "10.1007/JHEP06(2021)143",
    journal = "JHEP",
    volume = "06",
    pages = "143",
    year = "2021"
}

@article{Zeng:2019tlw,
    author = "Zeng, Yu-Pan and others",
    title = "{Probing quadruplet scalar dark matter at current and future $pp$ colliders}",
    eprint = "1910.09431",
    archivePrefix = "arXiv",
    primaryClass = "hep-ph",
    doi = "10.1103/PhysRevD.101.115033",
    journal = "Phys. Rev. D",
    volume = "101",
    number = "11",
    pages = "115033",
    year = "2020"
}

@article{Ostdiek:2015aga,
    author = "Ostdiek, Bryan",
    title = "{Constraining the minimal dark matter fiveplet with LHC searches}",
    eprint = "1506.03445",
    archivePrefix = "arXiv",
    primaryClass = "hep-ph",
    doi = "10.1103/PhysRevD.92.055008",
    journal = "Phys. Rev. D",
    volume = "92",
    pages = "055008",
    year = "2015"
}

@article{OHare:2020lva,
    author = "O'Hare, Ciaran A. J.",
    title = "{Can we overcome the neutrino floor at high masses?}",
    eprint = "2002.07499",
    archivePrefix = "arXiv",
    primaryClass = "astro-ph.CO",
    reportNumber = "CPPC-2020-13",
    doi = "10.1103/PhysRevD.102.063024",
    journal = "Phys. Rev. D",
    volume = "102",
    number = "6",
    pages = "063024",
    year = "2020"
}

@inproceedings{Baumann:2009ds,
    author = "Baumann, Daniel",
    title = "{Inflation}",
    booktitle = "{Theoretical Advanced Study Institute in Elementary Particle Physics}: {Physics of the Large and the Small}",
    eprint = "0907.5424",
    archivePrefix = "arXiv",
    primaryClass = "hep-th",
    reportNumber = "TASI-2009",
    doi = "10.1142/9789814327183_0010",
    pages = "523--686",
    year = "2011"
}

@article{Feiteira:2026qme,
    author = "Feiteira, Duarte and Lebedev, Oleg and Oliveira, Vin{\'\i}cius",
    title = "{Warm dark matter from freeze-in at stronger coupling}",
    eprint = "2602.20242",
    archivePrefix = "arXiv",
    primaryClass = "hep-ph",
    month = "2",
    year = "2026"
}

@article{Bernal:2025szh,
    author = "Bernal, Nicol{\'a}s and Neto, Jacinto P. and Silva-Malpartida, Javier and Queiroz, Farinaldo S.",
    title = "{Enabling thermal dark matter within the vanilla $L_\mu - L_\tau$ model}",
    eprint = "2507.02048",
    archivePrefix = "arXiv",
    primaryClass = "hep-ph",
    doi = "10.1103/8g85-c8sh",
    journal = "Phys. Rev. D",
    volume = "112",
    number = "7",
    pages = "075042",
    year = "2025"
}

@article{Bernal:2020bfj,
    author = {Bernal, Nicol{\'a}s and Rubio, Javier and Veerm{\"a}e, Hardi},
    title = "{Boosting Ultraviolet Freeze-in in NO Models}",
    eprint = "2004.13706",
    archivePrefix = "arXiv",
    primaryClass = "hep-ph",
    reportNumber = "PI/UAN-2020-670FT, HIP-2020-12/TH",
    doi = "10.1088/1475-7516/2020/06/047",
    journal = "JCAP",
    volume = "06",
    pages = "047",
    year = "2020"
}

@article{Yanagida:1979as,
    author = "Yanagida, Tsutomu",
    editor = "Sawada, Osamu and Sugamoto, Akio",
    title = "{Horizontal gauge symmetry and masses of neutrinos}",
    reportNumber = "KEK-79-18-95",
    journal = "Conf. Proc. C",
    volume = "7902131",
    pages = "95--99",
    year = "1979"
}

@article{Bernal:2020qyu,
    author = {Bernal, Nicol{\'a}s and Rubio, Javier and Veerm{\"a}e, Hardi},
    title = "{UV Freeze-in in Starobinsky Inflation}",
    eprint = "2006.02442",
    archivePrefix = "arXiv",
    primaryClass = "hep-ph",
    reportNumber = "PI/UAN-2020-673FT, HIP-2020-17/TH",
    doi = "10.1088/1475-7516/2020/10/021",
    journal = "JCAP",
    volume = "10",
    pages = "021",
    year = "2020"
}

@article{Garcia:2021iag,
    author = "García, Marcos A. G. and Kaneta, Kunio and Mambrini, Yann and Olive, Keith A. and Verner, Sarunas",
    title = "{Freeze-in from preheating}",
    eprint = "2109.13280",
    archivePrefix = "arXiv",
    primaryClass = "hep-ph",
    reportNumber = "UMN-TH-4101/21, FTPI-MINN-21/19, CERN-TH-2021-121",
    doi = "10.1088/1475-7516/2022/03/016",
    journal = "JCAP",
    volume = "03",
    number = "03",
    pages = "016",
    year = "2022"
}

@article{Chen:2023bwg,
    author = "Chen, Qing and Ding, Gui-Jun and Hill, Richard J.",
    title = "{General heavy WIMP nucleon elastic scattering}",
    eprint = "2309.02715",
    archivePrefix = "arXiv",
    primaryClass = "hep-ph",
    reportNumber = "USTC-ICTS/PCFT-23-26, FERMILAB-PUB-23-423-T",
    doi = "10.1103/PhysRevD.108.116023",
    journal = "Phys. Rev. D",
    volume = "108",
    number = "11",
    pages = "116023",
    year = "2023"
}

@article{Hisano:2004pv,
    author = "Hisano, Junji and Matsumoto, Shigeki and Nojiri, Mihoko M. and Saito, Osamu",
    title = "{Direct detection of the Wino and Higgsino-like neutralino dark matters at one-loop level}",
    eprint = "hep-ph/0407168",
    archivePrefix = "arXiv",
    reportNumber = "ICRR-REPORT-506-2004-4, YITP-04-39",
    doi = "10.1103/PhysRevD.71.015007",
    journal = "Phys. Rev. D",
    volume = "71",
    pages = "015007",
    year = "2005"
}

@phdthesis{Dutra:2019gqz,
    author = "Dutra, Maira",
    title = "{Origins for dark matter particles : from the ``WIMP miracle'' to the ``FIMP wonder''}",
    reportNumber = "tel-02100637, 2019SACLS059",
    school = "Orsay, LPT",
    year = "2019"
}

@article{Okada:2021uqk,
    author = "Okada, Nobuchika and Seto, Osamu",
    title = "{Superheavy WIMP dark matter from incomplete thermalization}",
    eprint = "2103.07832",
    archivePrefix = "arXiv",
    primaryClass = "hep-ph",
    reportNumber = "EPHOU-21-006",
    doi = "10.1016/j.physletb.2021.136528",
    journal = "Phys. Lett. B",
    volume = "820",
    pages = "136528",
    year = "2021"
}

@article{Bernal:2026clv,
    author = "Bernal, Nicol{\'a}s and Mukherjee, Sagnik and Unwin, James",
    title = "{Minimal Freeze-in Dark Matter: Reviving electroweak doublet dark matter with Boltzmann suppressed freeze-in}",
    eprint = "2602.10112",
    archivePrefix = "arXiv",
    primaryClass = "hep-ph",
    month = "2",
    year = "2026"
}

@article{LZ:2024zvo,
    author = "Aalbers, J. and others",
    collaboration = "LZ",
    title = "{Dark Matter Search Results from 4.2{\,}{\,}Tonne-Years of Exposure of the LUX-ZEPLIN (LZ) Experiment}",
    eprint = "2410.17036",
    archivePrefix = "arXiv",
    primaryClass = "hep-ex",
    reportNumber = "FERMILAB-PUB-24-0796-V",
    doi = "10.1103/4dyc-z8zf",
    journal = "Phys. Rev. Lett.",
    volume = "135",
    number = "1",
    pages = "011802",
    year = "2025"
}

@article{DARWIN:2016hyl,
    author = "Aalbers, J. and others",
    collaboration = "DARWIN",
    title = "{DARWIN: towards the ultimate dark matter detector}",
    eprint = "1606.07001",
    archivePrefix = "arXiv",
    primaryClass = "astro-ph.IM",
    doi = "10.1088/1475-7516/2016/11/017",
    journal = "JCAP",
    volume = "11",
    pages = "017",
    year = "2016"
}

@article{BICEP:2021xfz,
    author = "Ade, P. A. R. and others",
    collaboration = "BICEP, Keck",
    title = "{Improved Constraints on Primordial Gravitational Waves using Planck, WMAP, and BICEP/Keck Observations through the 2018 Observing Season}",
    eprint = "2110.00483",
    archivePrefix = "arXiv",
    primaryClass ="",
    doi = "10.1103/PhysRevLett.127.151301",
    journal = "Phys. Rev. Lett.",
    volume = "127",
    number = "15",
    pages = "151301",
    year = "2021"
}

@article{Giudice:2000ex,
    author = "Giudice, Gian Francesco and Kolb, Edward W. and Riotto, Antonio",
    title = "{Largest temperature of the radiation era and its cosmological implications}",
    eprint = "hep-ph/0005123",
    archivePrefix = "arXiv",
    reportNumber = "SNS-PH-00-05, FERMILAB-PUB-00-075-A, CERN-TH-2000-107",
    doi = "10.1103/PhysRevD.64.023508",
    journal = "Phys. Rev. D",
    volume = "64",
    pages = "023508",
    year = "2001"
}

@article{Aghaie:2025iyn,
    author = "Aghaie, Mohammad and Dondarini, Alessandro and Marino, Giulio and Panci, Paolo",
    title = "{Minimal Dark Matter in the sky: updated Indirect Detection probes}",
    eprint = "2507.17607",
    archivePrefix = "arXiv",
    primaryClass = "hep-ph",
    month = "7",
    year = "2025"
}

@article{Sarkar:1995dd,
    author = "Sarkar, Subir",
    title = "{Big bang nucleosynthesis and physics beyond the standard model}",
    eprint = "hep-ph/9602260",
    archivePrefix = "arXiv",
    reportNumber = "OUTP-95-16-P",
    doi = "10.1088/0034-4885/59/12/001",
    journal = "Rept. Prog. Phys.",
    volume = "59",
    pages = "1493--1610",
    year = "1996"
}

@article{Bernal:2019mhf,
    author = "Bernal, Nicol\'as and Elahi, Fatemeh and Maldonado, Carlos and Unwin, James",
    title = "{Ultraviolet Freeze-in and Non-Standard Cosmologies}",
    eprint = "1909.07992",
    archivePrefix = "arXiv",
    primaryClass ="",
    reportNumber = "PI/UAN-2019-654FT",
    doi = "10.1088/1475-7516/2019/11/026",
    journal = "JCAP",
    volume = "11",
    pages = "026",
    year = "2019"
}

@article{Barman:2021ugy,
    author = "Barman, Basabendu and Bernal, Nicol\'as",
    title = "{Gravitational SIMPs}",
    eprint = "2104.10699",
    archivePrefix = "arXiv",
    primaryClass ="",
    reportNumber = "PI/UAN-2021-688FT",
    doi = "10.1088/1475-7516/2021/06/011",
    journal = "JCAP",
    volume = "06",
    pages = "011",
    year = "2021"
}

@article{Barman:2022tzk,
    author = "Barman, Basabendu and Bernal, Nicol\'as and Xu, Yong and Zapata, {\'O}scar",
    title = "{Ultraviolet freeze-in with a time-dependent inflaton decay}",
    eprint = "2202.12906",
    archivePrefix = "arXiv",
    primaryClass ="",
    reportNumber = "PI/UAN-2022-710FT",
    doi = "10.1088/1475-7516/2022/07/019",
    journal = "JCAP",
    volume = "07",
    number = "07",
    pages = "019",
    year = "2022"
}

@article{Garcia:2020eof,
    author = "García, Marcos A. G. and Kaneta, Kunio and Mambrini, Yann and Olive, Keith A.",
    title = "{Reheating and Post-inflationary Production of Dark Matter}",
    eprint = "2004.08404",
    archivePrefix = "arXiv",
    primaryClass ="",
    reportNumber = "UMN--TH--3916/20, FTPI--MINN--20/06, IFT-UAM/CSIC-20-56",
    doi = "10.1103/PhysRevD.101.123507",
    journal = "Phys. Rev. D",
    volume = "101",
    number = "12",
    pages = "123507",
    year = "2020"
}

@article{Boddy:2024vgt,
    author = "Boddy, Kimberly K. and Freese, Katherine and Montefalcone, Gabriele and Shams Es Haghi, Barmak",
    title = "{Minimal dark matter freeze-in with low reheating temperatures and implications for direct detection}",
    eprint = "2405.06226",
    archivePrefix = "arXiv",
    primaryClass = "hep-ph",
    reportNumber = "UTWI-16-2024, NORDITA-2024-016",
    doi = "10.1103/PhysRevD.111.063537",
    journal = "Phys. Rev. D",
    volume = "111",
    number = "6",
    pages = "063537",
    year = "2025"
}

@article{Bernal:2024yhu,
    author = "Bernal, Nicol\'as and Deka, Kuldeep and Losada, Marta",
    title = "{Thermal dark matter with low-temperature reheating}",
    eprint = "2406.17039",
    archivePrefix = "arXiv",
    primaryClass ="",
    doi = "10.1088/1475-7516/2024/09/024",
    journal = "JCAP",
    volume = "09",
    pages = "024",
    year = "2024"
}

@article{Co:2020xaf,
    author = "Co, Raymond T. and González, Eric and Harigaya, Keisuke",
    title = "{Increasing Temperature toward the Completion of Reheating}",
    eprint = "2007.04328",
    archivePrefix = "arXiv",
    primaryClass ="",
    reportNumber = "LCTP-20-15",
    doi = "10.1088/1475-7516/2020/11/038",
    journal = "JCAP",
    volume = "11",
    pages = "038",
    year = "2020"
}

@article{Cosme:2024ndc,
    author = "Cosme, Catarina and Costa, Francesco and Lebedev, Oleg",
    title = "{Temperature evolution in the Early Universe and freeze-in at stronger coupling}",
    eprint = "2402.04743",
    archivePrefix = "arXiv",
    primaryClass ="",
    doi = "10.1088/1475-7516/2024/06/031",
    journal = "JCAP",
    volume = "06",
    pages = "031",
    year = "2024"
}

@article{Cosme:2023xpa,
    author = "Cosme, Catarina and Costa, Francesco and Lebedev, Oleg",
    title = "{Freeze-in at stronger coupling}",
    eprint = "2306.13061",
    archivePrefix = "arXiv",
    primaryClass ="",
    doi = "10.1103/PhysRevD.109.075038",
    journal = "Phys. Rev. D",
    volume = "109",
    number = "7",
    pages = "075038",
    year = "2024"
}

@article{Bernal:2025fdr,
    author = "Bernal, Nicol\'as and Deka, Kuldeep and Losada, Marta",
    title = "{Dark matter ultraviolet freeze-in in general reheating scenarios}",
    eprint = "2501.04774",
    archivePrefix = "arXiv",
    primaryClass ="",
    doi = "10.1103/PhysRevD.111.055034",
    journal = "Phys. Rev. D",
    volume = "111",
    number = "5",
    pages = "055034",
    year = "2025"
}

@article{Khan:2025keb,
    author = "Khan, Sarif and Kim, Jongkuk and Lee, Hyun Min",
    title = "{Higgs portal vector dark matter at a low reheating temperature}",
    eprint = "2503.17621",
    archivePrefix = "arXiv",
    primaryClass ="",
    doi = "10.1088/1475-7516/2025/06/040",
    journal = "JCAP",
    volume = "06",
    pages = "040",
    year = "2025"
}

@article{Arcadi:2024wwg,
    author = "Arcadi, Giorgio and Costa, Francesco and Goudelis, Andreas and Lebedev, Oleg",
    title = "{Higgs portal dark matter freeze-in at stronger coupling: observational benchmarks}",
    eprint = "2405.03760",
    archivePrefix = "arXiv",
    primaryClass ="",
    doi = "10.1007/JHEP07(2024)044",
    journal = "JHEP",
    volume = "07",
    pages = "044",
    year = "2024"
}

@article{Koivunen:2024vhr,
    author = "Koivunen, Niko and Lebedev, Oleg and Raidal, Martti",
    title = "{Probing sterile neutrino freeze-in at stronger coupling}",
    eprint = "2403.15533",
    archivePrefix = "arXiv",
    primaryClass ="",
    doi = "10.1140/epjc/s10052-024-13583-y",
    journal = "Eur. Phys. J. C",
    volume = "84",
    number = "11",
    pages = "1234",
    year = "2024"
}

@article{Bernal:2024ndy,
    author = "Bernal, Nicol{\'a}s and Fong, Chee Sheng and Zapata, {\'O}scar",
    title = "{Probing low-reheating scenarios with minimal freeze-in dark matter}",
    eprint = "2412.04550",
    archivePrefix = "arXiv",
    primaryClass ="",
    doi = "10.1007/JHEP02(2025)161",
    journal = "JHEP",
    volume = "02",
    pages = "161",
    year = "2025"
}

@article{Arcadi:2024obp,
    author = "Arcadi, Giorgio and Cabo-Almeida, David and Lebedev, Oleg",
    title = "{Z'-mediated dark matter freeze-in at stronger coupling}",
    eprint = "2409.02191",
    archivePrefix = "arXiv",
    primaryClass ="",
    doi = "10.1016/j.physletb.2025.139268",
    journal = "Phys. Lett. B",
    volume = "861",
    pages = "139268",
    year = "2025"
}

@article{Belanger:2024yoj,
    author = "B{\'e}langer, Genevi{\`e}ve and Bernal, Nicol{\'a}s and Pukhov, Alexander",
    title = "{Z'-mediated dark matter with low-temperature reheating}",
    eprint = "2412.12303",
    archivePrefix = "arXiv",
    primaryClass ="",
    doi = "10.1007/JHEP03(2025)079",
    journal = "JHEP",
    volume = "03",
    pages = "079",
    year = "2025"
}

@article{Hall:2009bx,
    author = "Hall, Lawrence J. and Jedamzik, Karsten and March-Russell, John and West, Stephen M.",
    title = "{Freeze-In Production of FIMP Dark Matter}",
    eprint = "0911.1120",
    archivePrefix = "arXiv",
    primaryClass ="",
    reportNumber = "OUTP-09-18-P, UCB-PTH-09-32",
    doi = "10.1007/JHEP03(2010)080",
    journal = "JHEP",
    volume = "03",
    pages = "080",
    year = "2010"
}

@article{Elahi:2014fsa,
    author = "Elahi, Fatemeh and Kolda, Christopher and Unwin, James",
    title = "{UltraViolet Freeze-in}",
    eprint = "1410.6157",
    archivePrefix = "arXiv",
    primaryClass ="",
    doi = "10.1007/JHEP03(2015)048",
    journal = "JHEP",
    volume = "03",
    pages = "048",
    year = "2015"
}

@article{Lee:2024wes,
    author = "Lee, Hyun Min and Park, Myeonghun and Sanz, Veronica",
    title = "{Gravity-Mediated Dark Matter at a low reheating temperature}",
    eprint = "2412.07850",
    archivePrefix = "arXiv",
    primaryClass ="",
    doi = "10.1007/JHEP05(2025)126",
    journal = "JHEP",
    volume = "05",
    pages = "126",
    year = "2025"
}

@article{Allahverdi:2020bys,
    author = "Allahverdi, Rouzbeh and others",
    title = "{The First Three Seconds: a Review of Possible Expansion Histories of the Early Universe}",
    eprint = "2006.16182",
    archivePrefix = "arXiv",
    primaryClass ="",
    reportNumber = "FERMILAB-PUB-20-242-A, KCL-PH-TH/2020-33, KEK-Cosmo-257,
  KEK-TH-2231, IPMU20-0070, PI/UAN-2020-674FT, RUP-20-22",
    doi = "10.21105/astro.2006.16182",
    journal = "Open J. Astrophys.",
    volume = "4",
    pages = "astro.2006.16182",
    year = "2021"
}

@article{Cirelli:2005uq,
    author = "Cirelli, Marco and Fornengo, Nicolao and Strumia, Alessandro",
    title = "{Minimal dark matter}",
    eprint = "hep-ph/0512090",
    archivePrefix = "arXiv",
    reportNumber = "DFTT40-2005, IFUP-TH-2005-34",
    doi = "10.1016/j.nuclphysb.2006.07.012",
    journal = "Nucl. Phys. B",
    volume = "753",
    pages = "178--194",
    year = "2006"
}

@article{Garcia:2017tuj,
    author = "García, Marcos A. G. and Mambrini, Yann and Olive, Keith A. and Peloso, Marco",
    title = "{Enhancement of the Dark Matter Abundance Before Reheating: Applications to Gravitino Dark Matter}",
    eprint = "1709.01549",
    archivePrefix = "arXiv",
    primaryClass ="",
    reportNumber = "LPT--ORSAY-17-36, UMN--TH--3635-17, FTPI--MINN--17-15",
    doi = "10.1103/PhysRevD.96.103510",
    journal = "Phys. Rev. D",
    volume = "96",
    number = "10",
    pages = "103510",
    year = "2017"
}

@article{Chowdhury:2023jft,
    author = "Chowdhury, Debtosh and Hait, Arpan",
    title = "{Thermalization in the presence of a time-dependent dissipation and its impact on dark matter production}",
    eprint = "2302.06654",
    archivePrefix = "arXiv",
    primaryClass ="",
    doi = "10.1007/JHEP09(2023)085",
    journal = "JHEP",
    volume = "09",
    pages = "085",
    year = "2023"
}

@article{Peccei:1977hh,
    author = "Peccei, R. D. and Quinn, Helen R.",
    title = "{CP Conservation in the Presence of Instantons}",
    reportNumber = "ITP-568-STANFORD",
    doi = "10.1103/PhysRevLett.38.1440",
    journal = "Phys. Rev. Lett.",
    volume = "38",
    pages = "1440--1443",
    year = "1977"
}

@article{XENON:2025vwd,
    author = "Aprile, E. and others",
    collaboration = "XENON",
    title = "{WIMP Dark Matter Search Using a 3.1 Tonne-Year Exposure of the XENONnT Experiment}",
    eprint = "2502.18005",
    archivePrefix = "arXiv",
    primaryClass = "hep-ex",
    doi = "10.1103/msw4-t342",
    journal = "Phys. Rev. Lett.",
    volume = "135",
    number = "22",
    pages = "221003",
    year = "2025"
}

@article{PandaX:2024qfu,
    author = "Bo, Zihao and others",
    collaboration = "PandaX",
    title = "{Dark Matter Search Results from 1.54{\,}{\,}Tonne{\textperiodcentered}Year Exposure of PandaX-4T}",
    eprint = "2408.00664",
    archivePrefix = "arXiv",
    primaryClass ="",
    doi = "10.1103/PhysRevLett.134.011805",
    journal = "Phys. Rev. Lett.",
    volume = "134",
    number = "1",
    pages = "011805",
    year = "2025"
}

@article{Bernal:2025qkj,
    author = "Bernal, Nicol{\'a}s and Cottin, Giovanna and D{\'\i}az S{\'a}ez, Basti{\'a}n and L{\'o}pez, Manuel",
    title = "{Testing frozen-in pNGB dark matter with a long-lived dark Higgs}",
    eprint = "2507.07089",
    archivePrefix = "arXiv",
    primaryClass = "hep-ph",
    doi = "10.1007/JHEP01(2026)081",
    journal = "JHEP",
    volume = "01",
    pages = "081",
    year = "2026"
}

@article{Bernal:2025osg,
    author = "Bernal, Nicol{\'a}s and Cervantes, Esau and Deka, Kuldeep and Hryczuk, Andrzej",
    title = "{Freezing-in cannibals with low-reheating temperature}",
    eprint = "2506.09155",
    archivePrefix = "arXiv",
    primaryClass ="",
    doi = "10.1007/JHEP09(2025)083",
    journal = "JHEP",
    volume = "09",
    pages = "083",
    year = "2025"
}

@article{Batell:2024dsi,
    author = "Batell, Brian and others",
    title = "{Conversations and deliberations: Non-standard cosmological epochs and expansion histories}",
    eprint = "2411.04780",
    archivePrefix = "arXiv",
    primaryClass ="",
    doi = "10.1142/S0217751X25300042",
    journal = "Int. J. Mod. Phys. A",
    volume = "40",
    number = "17",
    pages = "2530004",
    year = "2025"
}

@article{Bernal:2018qlk,
    author = "Bernal, Nicol{\'a}s and others",
    title = "{Spin-2 Portal Dark Matter}",
    eprint = "1803.01866",
    archivePrefix = "arXiv",
    primaryClass ="",
    doi = "10.1103/PhysRevD.97.115020",
    journal = "Phys. Rev. D",
    volume = "97",
    number = "11",
    pages = "115020",
    year = "2018"
}

@article{Bernal:2025fcl,
    author = "Bernal, Nicol{\'a}s and Mukherjee, Sagnik and Unwin, James",
    title = "{Boltzmann Suppressed Ultraviolet Freeze-in}",
    eprint = "2510.01311",
    archivePrefix = "arXiv",
    primaryClass ="",
    month = "10",
    year = "2025"
}

@article{Safdi:2025sfs,
    author = "Safdi, Benjamin R. and Xu, Weishuang Linda",
    title = "{Wino and Real Minimal Dark Matter Excluded by Fermi Gamma-Ray Observations}",
    eprint = "2507.15934",
    archivePrefix = "arXiv",
    primaryClass ="",
    reportNumber = "CERN-TH-2025-137",
    month = "7",
    year = "2025"
}

@article{Hisano:2004ds,
    author = "Hisano, Junji and Matsumoto, Shigeki. and Nojiri, Mihoko M. and Saito, Osamu",
    title = "{Non-perturbative effect on dark matter annihilation and gamma ray signature from galactic center}",
    eprint = "hep-ph/0412403",
    archivePrefix = "arXiv",
    reportNumber = "ICRR-REPORT-513-2004-11, YITP-04-73",
    doi = "10.1103/PhysRevD.71.063528",
    journal = "Phys. Rev. D",
    volume = "71",
    pages = "063528",
    year = "2005"
}

@article{Tucker-Smith:2001myb,
    author = "Tucker-Smith, David and Weiner, Neal",
    title = "{Inelastic dark matter}",
    eprint = "hep-ph/0101138",
    archivePrefix = "arXiv",
    reportNumber = "UCB-PTH-00-43, LBNL-47234, UW-PT-00-17",
    doi = "10.1103/PhysRevD.64.043502",
    journal = "Phys. Rev. D",
    volume = "64",
    pages = "043502",
    year = "2001"
}

@article{Essig:2007az,
    author = "Essig, Rouven",
    title = "{Direct Detection of Non-Chiral Dark Matter}",
    eprint = "0710.1668",
    archivePrefix = "arXiv",
    primaryClass ="",
    reportNumber = "RUNHETC-2007-20",
    doi = "10.1103/PhysRevD.78.015004",
    journal = "Phys. Rev. D",
    volume = "78",
    pages = "015004",
    year = "2008"
}

@article{LZ:2024psa,
    author = "Aalbers, J. and others",
    collaboration = "LZ",
    title = "{New constraints on ultraheavy dark matter from the LZ experiment}",
    eprint = "2402.08865",
    archivePrefix = "arXiv",
    primaryClass ="",
    reportNumber = "FERMILAB-PUB-24-0015-TD",
    doi = "10.1103/PhysRevD.109.112010",
    journal = "Phys. Rev. D",
    volume = "109",
    number = "11",
    pages = "112010",
    year = "2024"
}

\end{document}